\catcode`\@=11
\newif\if@fewtab\@fewtabtrue

{\count255=\time\divide\count255 by 60
\xdef\hourmin{\number\count255}
\multiply\count255 by-60\advance\count255 by\time
\xdef\hourmin{\hourmin:\ifnum\count255<10 0\fi\the\count255}}
\def\ps@draft{\let\@mkboth\@gobbletwo
    \def\@oddhead{}
    \def\@oddfoot
       {\hbox to 7 cm{$\scriptstyle Draft\ version:\ \draftdate$
       \hfil}\hskip -7cm\hfil\rm\thepage \hfil}
    \def\@evenhead{}\let\@evenfoot\@oddfoot}

\def\ceqno{\global\@fewtabfalse
    \ifcase\@eqcnt \def\@tempa{& & &}\or \def\@tempa{& &}
      \or \def\@tempa{&}
      \or\def\@tempa{}\fi\@tempa
{\rm(\theequation)}}

\def\aeqno#1{\global\@fewtabfalse
    \ifcase\@eqcnt \def\@tempa{& & &}\or \def\@tempa{& &}
      \or \def\@tempa{&}
      \or\def\@tempa{}\fi\@tempa
{\rm(\theequation,#1)}}

\def\label#1{\ifnum\draftcontrol=1
 \global\def\draftnote{$\scriptstyle #1$}\fi
 \@bsphack\if@filesw {\let\thepage\relax
   \def\protect{\noexpand\noexpand\noexpand}%
\xdef\@gtempa{\write\@auxout{\string
      \newlabel{#1}{{\@currentlabel}{\thepage}}}}}\@gtempa
   \if@nobreak \ifvmode\nobreak\fi\fi\fi
  \@esphack}

\def\alabel#1#2{\label{#1}\global\@fewtabfalse
    \ifcase\@eqcnt \def\@tempa{& & &}\or \def\@tempa{& &}
      \or \def\@tempa{&}
      \or\def\@tempa{}\fi\@tempa
{\hbox to 3cm{\phantom{\rm(\theequation,#2)}
\draftnote \hfil}\hskip -3cm {\rm(\theequation,#2)}}}

\def\clabel#1{\label{#1}\global\@fewtabfalse
    \ifcase\@eqcnt \def\@tempa{& & &}\or \def\@tempa{& &}
      \or \def\@tempa{&}
      \or\def\@tempa{}\fi\@tempa
{\hbox to 3cm{\phantom{\rm(\theequation)}
\draftnote \hfil}\hskip -3cm{\rm(\theequation)}}}

\def\eqnarray{\def\draftnote{{}}\global\@fewtabtrue
\stepcounter{equation}\let\@currentlabel=\theequation
\global\@eqnswtrue
\global\@eqcnt\z@\tabskip\@centering\let\\=\@eqncr
$$\halign to \displaywidth\bgroup\@eqnsel\hskip\@centering\@eqcnt\z@
  $\displaystyle\tabskip\z@{##}$&\global\@eqcnt\@ne
  \hskip 1\arraycolsep \hfil${##}$\hfil
  &\global\@eqcnt\tw@ \hskip 1\arraycolsep
$\displaystyle\tabskip\z@{##}$
\hfil  \tabskip\@centering&\global\@eqcnt\thr@@\llap{##}\tabskip\z@
\cr}

\def\endeqnarray{\@@eqncr\egroup
      \global\advance\c@equation\m@ne$$\global\@ignoretrue}

\def\@eqnnum{\hbox to 3cm{\phantom{\rm(\theequation)} \draftnote
                         \hfil}\hskip -3cm {\rm(\theequation)}}

\def\@@eqncr{\let\@tempa\relax
    \ifcase\@eqcnt \def\@tempa{& & &}\or \def\@tempa{& &}
      \or \def\@tempa{&}
      \or\def\@tempa{}
\fi\@tempa
\if@eqnsw
\if@fewtab\@eqnnum\fi
\stepcounter{equation}\fi\global
\@eqnswtrue\global\@eqcnt\z@\global\@fewtabtrue\cr}

\def\draftcite#1{\ifnum\draftcontrol=1#1\else{}\fi}

\def\@lbibitem[#1]#2{\item{}\hskip -3cm \hbox to 2cm
{\hfil$\scriptstyle\draftcite{#2}$}\hskip
1cm[\@biblabel{#1}]\if@filesw
     {\def\protect##1{\string ##1\space}\immediate
      \write\@auxout{\string\bibcite{#2}{#1}}}\fi\ignorespaces}

\def\@bibitem#1{\item\hskip -3cm \hbox to 2cm
{\hfil $\scriptstyle\draftcite{#1}$}\hskip 1cm
\if@filesw \immediate\write\@auxout
       {\string\bibcite{#1}{\the\value{\@listctr}}}\fi\ignorespaces}

 \def\nsection#1{\section{#1}\setcounter{equation}{0}}
     
     \def\nappendix#1{\vskip 1cm\no{\bf Appendix
         #1}\def\thesection{#1} \setcounter{equation}{0}}

\font\tendl=msbm10  scaled \magstep1
\font\sevendl=msbm7 scaled \magstep1
\font\fivedl=msbm5 scaled \magstep1
\font\tengl=eufm10  scaled \magstep1
\font\sevengl=eufm7 scaled \magstep1
\font\fivegl=eufm5 scaled \magstep1

\newfam\dlfam  
\textfont\dlfam=\tendl \scriptfont\dlfam=\sevendl
\scriptscriptfont\dlfam=\fivedl
\newfam\glfam  
\textfont\glfam=\tengl \scriptfont\glfam=\sevengl
\scriptscriptfont\glfam=\fivegl

\def\draftdate{\number\month/\number\day/\number\year\ \ \ \hourmin }

\global\def\draftcontrol{0}
\catcode`\@=12
\def\tilde{\widetilde}
\def\hat{\widehat}

\documentstyle[11pt,epsf]{article}

\def\theequation{{\thesection.\arabic{equation}}}

\newcommand{\be}{\begin{eqnarray}}
\newcommand{\en}{\end{eqnarray}\vs 0.5 cm}

\newcommand{\no}{\noindent}
\newcommand{\vs}{\vskip}

\newcommand{\un}{\underline}
\newcommand{\NR}{{{\bf R}}}

\newcommand{\NC}{{{\bf C}}}

\newcommand{\NZ}{{{\bf Z}}}

\newcommand{\Ng}{{\bf g}}
\newcommand{\Nh}{{\bf h}}
\newcommand{\qq}{\begin{eqnarray}}
\newcommand{\de}{\bar\partial}
\newcommand{\da}{\partial}
\newcommand{\ee}{{\rm e}}

\newcommand{\qqq}{\end{eqnarray}}

\newcommand{\tr}{\hbox{tr}}

\newcommand{\CA}{{\cal A}}
\newcommand{\CB}{{\cal B}}
\newcommand{\CC}{{\cal C}}

\newcommand{\CH}{{\cal H}}
\newcommand{\CI}{{\cal I}}

\newcommand{\CM}{{\cal M}}
\newcommand{\CN}{{\cal N}}
\newcommand{\CO}{{\cal O}}
\newcommand{\CP}{{\cal P}}

\newcommand{\CR}{{\cal R}}
\newcommand{\CS}{{\cal S}}

\newcommand{\CU}{{\cal U}}

\newcommand{\CZ}{{\cal Z}}

\newcommand{\m}{\hspace{0.025cm}}

\begin{document}
\title{Boundary \,WZW, $\,G/H$, $\,G/G\,$ and \,CS\, theories}
\author{\ 
\\Krzysztof Gaw\c{e}dzki \\ C.N.R.S., I.H.E.S.,
F-91440  Bures-sur-Yvette, France\\
and Laboratoire de Physique, ENS-Lyon,\\46, All\'ee d'Italie, F-69364 Lyon, 
France}
\date{ }
\maketitle

\vskip 0.3cm
\vskip 1 cm

\begin{abstract}
\vskip 0.3cm

\noindent  

\end{abstract}

We extend the analysis \cite{GTT} of the canonical 
structure of the Wess-Zumino-Witten theory 
to the bulk and boundary coset $G/H$ models. 
The phase spaces of the coset theories in the closed
and in the open geometry appear to coincide with those of 
a double Chern-Simons theory on two different
3-manifolds. In particular, we obtain an explicit description 
of the canonical structure of the boundary $G/G$ coset theory.
The latter may be easily quantized leading to an example
of a two-dimensional topological boundary field theory.
\vskip 1.3cm

\nsection{Introduction}
\vskip 0.2cm

Bidimensional boundary conformal field theory is 
a subject under intense study in view of its applications
to boundary phenomena in 1+1- or two-dimensional critical
systems and to the brane physics in string theory.
Although much progress has been achieved in understanding 
boundary CFT's since the seminal paper of Cardy \cite{Card}, 
much more remains to be done. The structure involved in the boundary 
theories is richer than in the bulk ones and their classification 
program involves new notions and an interphase with sophisticated
mathematics \cite{Moore1}\cite{PetZub}. One approach that 
offered a conceptual insight into the properties of correlation 
functions of boundary conformal models consisted of relating them 
to boundary states in three-dimensional topological field theories 
\cite{FFFS1}\cite{FFFS2}. In the simplest case of the boundary 
Wess-Zumino-Witten (WZW) models (conformal sigma models 
with a group $G$ as a target \cite{WittWZW}), the topological 
three-dimensional model appears to be the group $G$ 
Chern-Simons (CS) gauge theory \cite{Schwarz}\cite{WittCS}. In 
\cite{GTT} it has been shown how the relation between the boundary
WZW model and the CS theory arises in the canonical approach.
\vskip 0.3cm

The purpose of the present paper is to extend the analysis of \cite{GTT}
to the case of the coset $G/H$ models of conformal field theory
obtained by gauging in the group $G$ WZW model the adjoint action of 
a subgroup $H\subset G$. In the WZW model the simplest class of boundary 
conditions is obtained by restricting the boundary values of the 
classical $G$-valued field $g$ to fixed conjugacy classes in the group 
labeled by weights of the Lie algebra $\Ng$ of $\,G$. Such boundary 
conditions reduce to the Dirichlet conditions for toroidal targets.
It was shown in \cite{GTT} that the phase space of the WZW model
on a strip with such boundary conditions is isomorphic to the phase 
space of the CS theory on the time-line $\NR$ times a disc $D$ with 
two time-like Wilson lines.
In the coset models we shall use more general boundary conditions 
requiring that field $g$ belongs on the boundary components to pointwise
product of group $G$ and subgroup $H$ conjugacy classes. 
The phase space of the coset theory on a strip with such boundary
conditions becomes isomorphic to the phase space of the double
CS theory on $\NR\times D$ with group $G$ and group $H$ gauge
fields, both coupled to two time-like Wilson lines. In particular, 
the phase space of the boundary $G/G$ coset model\footnote{There
are other ways to impose boundary conditions in the $G/G$ theory
\cite{FGbc} that permit to relate it to the boundary topological 
Poisson sigma models of \cite{CatFel}} becomes isomorphic to 
the moduli space of flat connections on the 2-sphere with four 
punctures. The latter case lends itself easily to quantization giving 
rise to an example of the two-dimensional boundary topological field 
theory, a structure that promises to play the role of the $K$-theory 
of loop spaces \cite{Moore1}.
\vskip 0.3cm

Much of the motivation for the present work stemmed from interaction
with Volker Schomerus who generously shared his insights with the author.
The discussions with Laurent Freidel are also greatfully acknowledged.
Special thanks are due to the Erwin Schr\"{o}dinger Institute in Vienna 
were this work was started.
\vskip 0.9cm

\nsection{Action functionals of the WZW and coset theories}
\vskip 0.2cm

The Wess-Zumino-Witten model is a specific two-dimensional sigma 
model with a group manifold $G$ as the target. For simplicity,
we shall assume that $G$ is compact connected and simply connected. 
We shall denote by $\Ng$ the Lie algebra of $G$. The $G$-valued fields 
of the WZW model are defined on two-dimensional surfaces $\Sigma$ 
(the ``worldsheets'') that we shall take oriented and equipped with 
a conformal or pseudo-conformal structure. The action of the model 
in the Euclidean signature is the sum of two terms: 
\qq
S(g)\ =\ {_k\over^{4\pi i}}\int\limits_\Sigma\tr\,(g^{-1}\da g)  
(g^{-1}\de g)\ +\ S^{WZ}(g)\,.
\label{action}
\qqq
Above, $\,\tr\,$ stands for the properly normalized Killing form.
The second (Wess-Zumino) term in the action is related to the
canonical closed 3-form $\,\chi(g)={1\over3}\,\tr\,(g^{-1}dg)^3\,$
on $G$. \,Informally, it may be written as
\qq
S^{WZ}(g)\ =\ {_k\over^{4\pi i}}\int\limits_\Sigma g^* \omega
\label{inf}
\qqq
where $\omega$ is a 2-form on $G$ such that $d\omega=\chi$.
This definition is, however, problematic since there is no global 
$\omega$ with the last property. If $\Sigma$ has no boundary then 
the problem may be solved by setting \cite{WittWZW} 
\qq
S^{WZ}(g)\ =\ {_k\over^{4\pi i}}\int\limits_B{\tilde g}^*\chi\,,
\label{wz}
\qqq
where $B$ is a 3-manifold such that $\da B=\Sigma$ and $\tilde g$
extends $g$ to $B$. It is well known that this determines $S^{WZ}(g)$
modulo $2\pi i k$ so that the amplitudes $\exp[-S^{WZ}(g)]$ are well 
defined if $k$ is integer. By the Stokes Theorem, the definition 
(\ref{wz}) reduces to the naive expression (\ref{inf}) whenever $g$ 
maps into the domain 
of a local form $\omega$. The variation $\delta S^{WZ}(g)$ 
involves only the 3-form $\chi$ so that the classical equations are 
determined unambiguously. 
\vskip 0.4cm

For surfaces with boundary, one should impose proper boundary
conditions on fields $g$. Let $\da\Sigma=\sqcup S^1_n$ where 
$S^1_n$ are disjoint circles always considered with the orientation
inherited from $\Sigma$. Let, for $\,\mu\,$ in the Cartan
subalgebra of $\,\Ng$, $\,C_\mu$ denote 
the conjugacy class $\,\{\gamma\,\ee^{2\pi i\mu}\gamma^{-1}\,|\,
\gamma\in G\}$. \,We shall require that 
\qq
g(S^1_n)\ \subset\ C_{\mu_n}\,.
\label{cbc}
\qqq
These are the so called fully symmetric conformal boundary conditions.
When restricted to a conjugacy class $C_\mu$, the 3-form $\chi$ becomes 
exact. In particular, the 2-form  
\qq
\omega_\mu(g)\ =\ \tr\,(\gamma^{-1}d\gamma)\,\ee^{2\pi i\mu}(\gamma^{-1}
d\gamma)\,\ee^{-2\pi i\mu}
\qqq
on $\m C_\mu$ satisfies $\m d\omega_\mu=\chi|_{C_\mu}$. \,Let $\,\Sigma'
=\Sigma\#(\sqcup D_n)\,$ be the surface without boundary obtained
from $\Sigma$ by gluing discs $D_n$ to the boundary components 
$S^1_n$ of $\m\Sigma$, see Fig.\,\,1.

\leavevmode\epsffile[-5 -20 345 195]{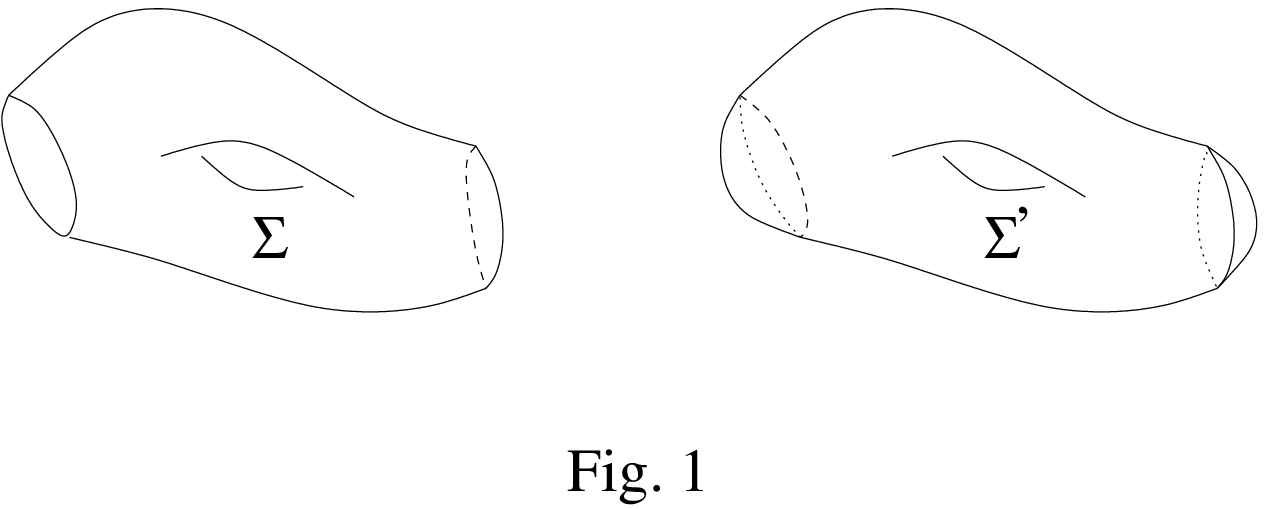}

\noindent Each field $g$ satisfying the boundary conditions (\ref{cbc}) 
may be extended to $\,g':\Sigma'\rightarrow G\,$ in such a way that
$\,g'(D_n)\subset C_{\mu_n}\,$ (the conjugacy classes are
simply connected). Following \cite{ASchom}\cite{ist}, we shall define 
the WZ-action of the field $g$ satisfying the boundary conditions 
(\ref{cbc}) by setting
\qq
S^{WZ}(g)\ =\ S^{WZ }(g')\ -\ {_k\over^{4\pi i}}\sum\limits_n
\int\limits_{D_n}{g'}^*\omega_{\mu_n}\,.
\label{bwz}
\qqq
This again reduces to the naive definition whenever $g$ maps into the 
domain of a form $\omega$ such that $\m d\omega=\chi$, provided that the 
restrictions of $\m\omega$ to $\m C_{\mu_n}$ coincide with $\m\omega_{\mu_n}$. 
A different choice of the restrictions would change the boundary 
contributions to the classical equations. As explained in \cite{ist}, 
(for $k\not=0$) the right hand side of (\ref{bwz}) is well defined 
modulo $2\pi i$ iff $k$ is an integer and $C_{\mu_n}=C_{\lambda_n/k}$ 
for integrable 
weights\footnote{The weights integrable at level $k$ are the ones
lying in the positive Weyl alcove inflated by $k$} $\lambda_n$. 
The boundary conditions are thus labeled 
by the same set as the bulk primary fields of the current algebra
also corresponding to integrable weights. This is an illustration 
of the Cardy's theory of boundary conditions \cite{Card}.
\vskip 0.4cm

The $G/H$ coset theories \cite{GKO} may be realized as the versions of
the group $G$ WZW theory where the adjoint action of the subgroup
$H\subset G$ has been gauged \cite{BarRS}\cite{GK1}\cite{KPS}\cite{GK2}. 
Let $A$ denote a
1-form with values in $i\Nh$, where $\Nh$ stands for the Lie algebra of $H$.
For $\da\Sigma=\emptyset$, the action of the theory coupled to the gauge 
field $A=A^{10}+\m A^{01}$ is
\qq
S(g,A)\ =\ S(g)\ -\ {_k\over^{2\pi i}}\int\limits_\Sigma\tr\,\Big[\,
(g\da g^{-1})A^{01}+A^{10}(g^{-1}\de g)+gA^{10}g^{-1}\hspace{-0.08cm}
A^{01}-A^{10}A^{01}\Big].
\label{cgf}
\qqq
In fact, getting rid of the so called ``fixed point problem'' 
\cite{Schell}\cite{FSSchw} (that obstructs factorization properties 
of the theory) requires considering the WZW theory coupled 
to gauge fields in non-trivial $H/Z$-bundles, where $Z$ is the 
intersection of the center of $G$ with $H$ \cite{MooSei}\cite{Hori}. 
For simplicity, we shall not do it here. For the surfaces with 
boundary $\da\Sigma=\sqcup S^1_n$, we shall use the same formula 
(\ref{cgf}) to include the coupling to the gauge field, but we shall 
admit more general boundary conditions for the field $g$ than 
the ones considered before. Namely, we shall assume that
\qq
g|_{S^1_n}\,=\ g_n h_n^{-1}\qquad {\rm with}\quad g_n:S^1_n\rightarrow 
C^G_{\mu_n},\quad h_n:S^1_n\rightarrow C^H_{\mu_n}\,.
\label{ghbc}
\qqq
In other words, we shall admit fields $g$ that, on each boundary component,
are a pointwise product of loops in conjugacy classes of, respectively,
group $G$ and group $H$, keeping also track of the decomposition
factors\footnote{The 
decomposition of elements of the pointwise product $\,C^G_{\mu_n}
(C^H_{\nu_n})^{-1}$ might not be unique.}. We shall label such conditions
by pairs $\,(\mu_n,\nu_n)\equiv M_n$. For $\nu_n=0$, they reduce to
the conditions considered in the previous section. We still need to 
generalize the definition of the Wess-Zumino term of the action
to fields $g$ satisfying (\ref{ghbc}). Such fields may be extended to 
maps $\,g':\Sigma'\rightarrow G\,$ in such a way that 
\qq
&g'|_{D_n}=g'_n{h'_n}^{\hspace{-0.08cm}-1}\qquad{\rm with}
\qquad g'|_{S^1_n}=g_n\,,\quad\,h'_n|_{S^1_n}=h_n&\cr\cr 
&{\rm and}\qquad g'_n(D_n)\subset C^G_{\mu_n}\,,\qquad\,h'_n(D_n)\subset 
C^H_{\nu_n}\,.&
\nonumber
\qqq
We shall define then
\qq
S^{WZ}(g)\,=\, S^{WZ}(g')\,-\,{_k\over^{4\pi i}}\sum\limits_n\int\limits_{D_n}
\Big[\,{g'_n}^*\omega^G_{\mu_n}\,-\,{h'_n}^*\omega^H_{\nu_n}\,+\,\tr\,
({g'_n}^{\hspace{-0.08cm}-1}dg'_n)({h'_n}^{\hspace{-0.08cm}-1}dh'_n)\,\Big]\,.
\label{ghac}
\qqq
The form in the brackets has $\,(g'_n{h'_n}^{\hspace{-0.08cm}-1})^*\chi\,$ 
as the exterior derivative which assures invariance of the right hand side
under continuous deformations of $\m g'_n$ and $\m h'_n$ inside discs
$\m D_n$. \,If $k\not=0$ and $H$ is simply connected then a slight 
extension of the argument in \cite{ist} shows that $S^{WZ}(g)$ given 
by (\ref{ghbc}) is well 
defined modulo $2\pi i$ iff $k$ is integer and $\,C^G_{\mu_n}
=C^G_{\lambda_n/k}$, $\,C^H_{\nu_n}=C^H_{\eta_n/k}$, where $\lambda_n$
and $\eta_n$ are integrable weights of $\Ng$ and $\Nh$,
respectively. We shall use (\ref{action}), (\ref{ghac}) and (\ref{cgf}) 
to define the complete gauged action $S(g,A)$. With the above choices of
the boundary labels, the gauge invariance 
\qq
\exp[-S(hgh^{-1},hAh^{-1}\hspace{-0.08cm}+hdh^{-1})]\ =\ \exp[-S(g,A)]
\label{ginv}
\qqq
holds for $h:\Sigma\rightarrow H$. If $H$ has an abelian factor, then 
the same selection of boundary conditions is imposed if we add to 
the demand that $\exp[-S^{WZ}(g)]$ be well defined the requirement 
of the gauge-invariance (\ref{ginv}). For example, for the parafermionic 
$SU(2)/U(1)$ coset theory, this restricts the boundary labels to pairs 
$\,(\lambda_n,\eta_n)=(j_n\sigma_3,\m m_n\sigma_3)\,$ with $j_n=0,{1\over2},
\dots,{k\over2}$ and $m_n=0,{1\over2},\dots,{k}-{1\over2}$. 
The labels of 
the parafermionic primary states $(j,m)$ have additional selection rule 
$j=m\ {\rm mod}\,1$ and the identification $(j,m)\sim({k\over2}-j,m+
{k\over2}\ {\rm mod}\,k)$. For the boundary labels, the first may be 
imposed by requiring the gauge invariance with respect to 
$\,h:\Sigma\rightarrow U(1)/\NZ_2\,$ and the second by identifying 
the decompositions $g_nh_n^{-1}$ and $(-g_n)(-h_n)^{-1}$. Similarly, 
in the general case we may impose the local $\,H/Z\,$ gauge invariance 
and identify the decompositions differing by an element in 
$Z$ \cite{inprep}. Such restrictions lead to the same labeling of 
the boundary conditions and of the primary fields, but is not obligatory 
if we ignore the fixed point problem.
\vskip 0.4cm

Since the gauge field $A$ enters quadratically into the action
(\ref{cgf}), it may be eliminated classically (and also quantum
mechanically) from the equations of motion. What results is  
a sigma model with the space $\,G/Ad(H)\,$  of the orbits
of the adjoint action of $\,H\,$ on $\,G\,$ as the target.
The target space $\,G/Ad(H)\,$ (that may be singular) comes equipped 
with a specific metric, a non-meric volume form (``dilaton field'') 
and a 2-form. Let $\,[g]\,$ denote the projection of $\,g\,$ to 
$\,G/Ad(H)$. \,The boundary conditions (\ref{ghbc}) restrict
the boundary values of $\,[g]\,$ to the projection to
$\,G/Ad(H)\,$ of the rotated $G$-group conjugacy class 
$\,C^G_{_{\mu_n}}\ee^{-2\pi i\m\nu_n}$ (but contain more data
if the decomposition $\,g_nh_n^{-1}\m$ is not unique).
For example, for $\,G=SU(2)\,$ with elements $\,(\matrix{_z&_{-z'}
\cr^{{\bar z}'}&^{\bar z}})$, \m where $\,|z|^2+|z'|^2=1$, \m and for 
$\,H=U(1)\m$, the coset space $\,G/Ad(H)\,$ may be identified with 
the unit disc $\,D=\{\m z\,|\,|z|\leq1\m\}$. \,The boundary conditions 
(\ref{ghbc}) with $\,(\lambda_n,\nu_n)\,$ corresponding to $\,(j_n,m_n)\m$
restrict the boundary values of $\,[g]\,$ to the intervals
$\,[\ee^{\m2\pi i(j-m)/k},\,\ee^{-2\pi i(j+m)/k}]\subset D\,$ 
with $2k$ end-points on the disc boundary. Since the conjugacy 
classes of $\,U(1)\,$ are composed of single points, the decomposition 
$\,g_nh_n^{-1}\m$ is unique in this case, given the conjugacy class
labels. Imposing the restriction $\,j=m\ {\rm mod}\,1\,$ eliminates 
half of the interval endpoints \cite{MMS}. 
\vskip 0.9cm

\nsection{Canonical structure of the WZW and coset theories}
\vskip 0.2cm

The classical field theory studies the solutions of the
variational problem $\delta S=0$ determined by the action
functional $S$. The space of solutions on a worldsheet
with the product structure $\NR\times\CN$ and Minkowski signature
admits a canonical closed 2-form $\Omega$, see e.g.\,\,\cite{COQG} 
or \cite{GTT}. If the latter is degenerate (a situation in gauge theories, 
where the degenerate directions correspond to local gauge transformations), 
one passes to the space of leaves of the degeneration distribution. 
By definition, the resulting space is the phase space of the theory 
and it carries the canonical symplectic structure\footnote{We ignore 
here the eventual problems with the infinite-dimensional character 
of the spaces and singularities that may be usually dealt with in 
concrete situations}.
\vskip 0.2cm

\subsection{Bulk WZW model}

Let us start with the well known case of the WZW model on the cylinder 
$\,\Sigma=\NR\times S^1=\{(t,\,x\,\,{\rm mod}\,2\pi)\}$. \,The 
variational equation $\delta S=0$ becomes a non-linear version of the 
wave equation
\qq
\da_+(g^{-1}\da_-g)\ =\ 0\,,
\label{cewz}
\qqq
where $\da_\pm=\da_{x^\pm}$ with $x^\pm=x\pm t$. The solutions may be 
labeled by the Cauchy data $\,g(t,\,\cdot\,)\,$ and $\,(g^{-1}\da_tg)(t,
\,\cdot\,)$. \,The space of solutions forms the phase space 
$\,\CP^{WZW}$ of the bulk WZW model. Its canonical symplectic 
form is given by the expression \cite{COQG}
\qq
\Omega^{WZ}\ =\ {_k\over^{4\pi}}\int\limits_0^{2\pi}\tr\,\Big[
-\delta(g^{-1}\da_tg)\,\,g^{-1}\delta g\ +\ 2\,(g^{-1}\da_+g)\,
(g^{-1}\delta g)^2\Big](t,x)\,\,dx\,\,
\label{owz}
\qqq
which is $t$-independent\footnote{We use the symbol $\,\delta\,$ for 
the exterior derivative on the space of classical solutions}. Similarly 
as for the wave equation, the general solution of (\ref{cewz}) may 
be decomposed as
\qq
g(t,x)\ =\ g_\ell(x^+)\,g_r(x^-)^{-1}\,.
\label{lrm}
\qqq
The left-right movers $\,g_{\ell,r}:\NR\rightarrow G\,$ are not
necessarily periodic but satisfy $\,g_{\ell,r}(y+2\pi)=g_{\ell,r}(y)
\,\gamma\,$ for the same $\gamma\in G$.  They are determined uniquely 
up to the simultaneous right multiplication by an element of $G$. 
The expression of the symplectic form in terms of the left-right 
movers is described in Appendix A. \,The currents
\qq
J_\ell\ =\ ik\, g\m\partial_+g^{-1}\ =\ ik\, g_\ell\m\partial_+g_\ell^{-1}\,,
\qquad J_r\ =\ ik\, g^{-1}\partial_-g\ =\ ik\, g_r\m\partial_+g_r^{-1}\,
\label{curr}
\qqq
generate the current algebra symmetries of the theory. The
conformal symmetries are generated by the components 
\qq
T_\ell\ =\ {_1\over^{2k}}\,\tr\,J_\ell^2\,,\qquad
T_r\ =\ {_1\over^{2k}}\,\tr\,J_r^2\,
\label{emt}
\qqq
of the energy momentum tensor.
\vskip 0.2cm

\subsection{Bulk $G/H$ model}

In the same cylindrical worldsheet geometry $\Sigma=\NR\times S^1$,
the classical equations for the coset $G/H$ model take the form:
\qq
D_+(g^{-1}D_-g)\ =\ 0\,,\quad E\,\,g^{-1}D_-g\ =\ 0\ =\ E\,\,g\,D_+g^{-1}\,,
\quad F(A)\ =\ 0\,,
\label{cegh}
\qqq
where $\,D_\pm=\da_\pm+[A_\pm,\,\cdot\,]\,$ are the light-cone covariant
derivatives, $\,E$ is the orthogonal projection of $\Ng$ onto
$\Nh$ and $F(A)=dA+A^2$ is the curvature of $A$. The equations
are preserved by the $H$-valued gauge transformations of the fields. 
The gauge transformations provide for the degeneration of the canonical
closed 2-form on the space of solutions so that the phase space
$\,\CP^{G/H}$ of the bulk coset theory is composed of the gauge-orbits
of the solutions of the classical equations (\ref{cegh}). 
\vskip 0.3cm

The gauge-orbits of solutions may be parametrized in a more effective 
way. The flat gauge field $A$ may be expressed as $\,h^{-1}dh\,$ 
for $\,h:\NR^2\rightarrow H\,$ such that $\,h(t,x+2\pi)
=\rho^{-1}h(t,x)\,$ for some $\rho\in H$. The map $h$ is determined 
uniquely up to the left multiplication by an element of $H$. 
\,Let us set $\,\tilde g=hgh^{-1}$. Note that $\,\tilde 
g:\NR^2\rightarrow G\,$ with $\,\tilde g(t,y+2\pi)=\rho^{-1}\tilde g(t,y)\, 
\rho$. \,In terms of field $\,\tilde g$, \,the classical equations 
reduce to
\qq
\da_+(\tilde g^{-1}\da_-\tilde g)\ =\ 0\,,\quad E\,\,\tilde g^{-1}
\da_-\tilde g\ =\ 0\ =\ E\,\,\tilde g\,\da_+{\tilde g}^{-1}\,.
\label{ceghp}
\qqq
The gauge-orbits of the classical solutions of (\ref{cegh}) are in
ono-to-one correspondence with the orbits of pairs 
$\,(\tilde g,\rho)\,$ under the simultaneous conjugation by elements 
of $H$. In terms of these data, the canonical symplectic form on 
$\,\CP^{G/H}$, obtained following the general prescriptions
of \cite{COQG}, is given by
\qq
&&\Omega^{G/H}\ =\ {_k\over^{4\pi}}\int\limits_0^{2\pi}\tr\,\Big[
-\delta({\tilde g}^{-1}\da_t\tilde g)\,\,{\tilde g}^{-1}\delta 
\tilde g\ +\ 2\,({\tilde g}^{-1}\da_+\tilde g)\,
({\tilde g}^{-1}\delta\tilde g)^2\Big](t,x)\,\,dx\cr
&&+\,{_k\over^{4\pi}}\,\tr\,\Big[\,(\delta\rho)\rho^{-1}\Big(\tilde g
(t,0)^{-1}(\delta\rho)\rho^{-1}\tilde g(t,0)\,-\,({\tilde g}^{-1}
\delta\tilde g)(t,0)-((\delta\tilde g){\tilde g}^{-1})(t,0)\Big)\,\Big]\,\,
\label{GHsf0}
\qqq
for any fixed $t$. \,The solutions of the classical equations
(\ref{ceghp}) may be expressed again by the left-right movers:
$\,\tilde g(t,x)=g_\ell(x^+)\m g_r(x^-)^{-1}\m$, \,where  
$\,g_{\ell,r}:\NR\rightarrow G\,$ are such that
\qq
E\,\,g_{\ell,r}\da_yg_{\ell,r}^{-1}\ =\ 0\qquad{\rm and}
\qquad g_{\ell,r}(y+2\pi)\ =\ \rho^{-1}\m g_{\ell,r}(y)\,\gamma\,.
\label{cond}
\qqq
Given $\,\tilde g\m$, \,the one-dimensional fields $\,g_{\ell,r}$ are 
determined up to the simultaneous right multiplication by an element of $G$. 
The expression for the symplectic form $\,\Omega^{G/H}\m$ in terms of 
the left-right movers is given in Appendix A. The left-right components
of the energy-momentum tensor
\qq
&&T_\ell\ =\ -\m{_k\over^2}\,\tr\,(g\m D_+g^{-1})^2\ =\ 
-\m{_k\over^2}\,\tr\,(g_\ell\m\partial_+g_\ell^{-1})^2\,,\cr
\ \label{GHemt}\\
&&T_r\ =\ -\m{_k\over^2}\,\tr\,(g^{-1}\m D_-g)^2\ =\ -\m{_k
\over^2}\,\tr\,(g_r\m\partial_+g_r^{-1})^2\,
\nonumber
\qqq
generate the conformal symmetries of the bulk coset model.
\vskip 0.2cm

\subsection{Bulk $G/G$ model}

For the topological coset $G/G$ theory, the classical equations 
(\ref{ceghp}) reduce to $\,{\tilde g}^{-1}d\tilde g=0\m$, 
\,i.e.\ $\tilde g\,$ is constant and it commutes with the monodromy
$\,\rho$. \,The phase space $\,\CP^{G/G}$ may be identified with 
the space of commuting pairs $(\tilde g,\rho)$ in $G$ modulo 
simultaneous conjugations. It is finite-dimensional, in agreement with the
topological character of the theory. It comes equipped with the symplectic 
form
\qq
\Omega^{G/G}\ =\ 
{_k\over^{4\pi}}\,\tr\,\Big[\,(\delta\rho)\rho^{-1}\,\Big(\tilde g^{-1}
(\delta\rho)\rho^{-1}\tilde g\,-\,{\tilde g}^{-1}\delta\tilde g
-(\delta\tilde g){\tilde g}^{-1}\Big)\,\Big]\,.
\label{GGsf}
\qqq
Up to a simultaneous conjugation, $\,\tilde g=\ee^{\m2\pi i\,\mu}\,$
and $\,\rho=\ee^{\m2\pi i\,\nu}\,$ for $\,\mu\,$ and $\,\nu\,$
in the Cartan algebra and the symplectic form becomes a
a constant form on the product of two copies of the Cartan algebra.
\qq
\Omega^{G/G}\ =\ 2\pi\m k\m\,\tr\,[\m d\nu\,d\mu\m]\,.
\label{GGsf1}
\qqq
In particular, conjugation-invariant functions of $\,\tilde g\,$
Poisson-commute and so do those of $\,\rho$.
\vskip 0.2cm

\subsection{Boundary \,WZW\, model}

The canonical treatment of the boundary theories is quite analogous 
to that of the bulk ones, except for the necessity to treat the boundary 
contributions. We consider the strip geometry $\Sigma=\NR\times[0,\pi]$ 
with Minkowski signature and impose on the field $\,g:\Sigma
\rightarrow G\,$ of the WZW model the boundary conditions
discussed in Sect.\,\,2\m:
\qq
g(t,0)\,\in\,C_{\mu_0}\,,\quad\qquad g(t,\pi)\,\in\,C_{\mu_\pi}\,.
\label{bc12}
\qqq
For variations $\delta g$ tangent to the space of fields respecting 
conditions (\ref{bc12}), the classical equations $\,\delta S(g)=0\,$ 
reduce to the bulk equation (\ref{cewz}) supplemented with the boundary
equations
\qq
g^{-1}\da_-g\ +\ g\,\da_+g^{-1}\ =\ 0\qquad {\rm for}\ \,x=0,\,\pi\,.
\label{bewz}
\qqq
The classical solutions obeying (\ref{bc12}) form the phase space 
$\,\CP^{WZ}_{_{\mu_0\mu_\pi}}\,$ of the boundary WZW model. Its 
symplectic form is given by \cite{GTT}
\qq
\Omega^{WZ}_{\mu_0\mu_\pi}
\ \ =\ \ {_k\over^{4\pi}}\int\limits_0^\pi\tr\,\Big[-\delta(g^{-1}\da_tg)\,
g^{-1}\delta g\ +\ 2(g^{-1}\da_+g)\,(g^{-1}\delta g)^2\Big](t,x)\,\, dx
\,\,\cr
+\ {_k\over^{4\pi}}\,\Big[\,\omega_{\mu_0}(g(t,0))\ 
-\ \omega_{\mu_\pi}(g(t,\pi))\,\Big]\,\,
\label{obd}
\qqq
for any fixed $t$. \,As in the bulk, the classical equations may be 
solved explicitly, as was described in \cite{GTT}. We have\footnote{We 
use here a slightly different parametrization of the solutions than in
\cite{GTT}.} 
\qq
g(t,x)\ =\ g_\ell(x^+)\,\,m_0\,\,g_\ell(-x^-)^{-1}\ =\ 
g_\ell(x^+)\,\,m_\pi\,\,g_\ell(2\pi-x^-)^{-1}\,,
\label{wzcs}
\qqq
where $\,m_0\in C_{\mu_0}$, $\,m_\pi\in C_{\mu_\pi}$
and $\,g_\ell:\NR\rightarrow G\,$ satisfy
\qq
g_\ell(y+2\pi)\ =\ g_\ell(y)\,\gamma\qquad {\rm for}\qquad 
\gamma=m_0^{-1} m_\pi\,.
\label{wzm} 
\qqq
Note that the boundary conditions (\ref{bc12}) are fulfilled.
The orbits of $\,(g_\ell,m_0,m_\pi)\,$ under the right multiplication 
of $g_\ell$ by elements of $G$ accompanied by the inverse adjoint action 
on $m_0$ and $m_\pi$ are in one-to-one correspondence
with the classical solutions. The expression of the symplectic form 
in terms of these data is given in Appendix A. \,The boundary WZW theory
has a single current $\,J=ik\, g_\ell\partial_+g_\ell^{-1}\,$ with the
corresponding energy-momentum tensor $\,T={1\over{2k}}\,\tr\,J^2$.
\vskip 0.2cm

\subsection{Boundary $\,G/H\,$ model}

For the boundary coset $G/H$ model with the $G$-valued field $g$ and 
$i\Nh$-valued gauge-field $A$ defined on the strip $\,\NR\times[0,\pi]$, 
\,we shall impose the boundary conditions
\qq
g(t,0)\ =\ g_0(t)\,h_0(t)^{-1}\,,\quad\qquad g(t,\pi)\ =\ g_\pi(t)\,
h_\pi(t)^{-1}
\label{GHbc12}
\qqq
with $\,g_0,\ h_0,\ g_\pi$ \,and $\,h_\pi\,$ mapping the boundary
lines into the conjugacy classes $\,C^G_{\mu_0}$, $\,C^H_{\nu_0}$, 
$\,C^G_{\mu_\pi}$ \,and $\,C^H_{\nu_\pi}$, \,respectively, see 
(\ref{ghbc}). The gauge fields $A$ will not be restricted.
We shall label such boundary conditions by the pairs $(M_0,M_\pi)$, 
where $M_0=(\mu_0,\nu_0)$ and $M_\pi=(\mu_\pi,\nu_\pi)$. \,The variational 
equations $\,\delta S(g,A)=0\,$ reduce now to the bulk equations 
(\ref{cegh}) supplemented with the boundary equations
\qq
&h_0^{-1}D_t\,h_0\ \ =\ \ 0\ \ =\ \ h_\pi^{-1}D_t\,h_\pi\,,&\label{bde1}
\\ \cr
&(g^{-1}D_-g)(\,\cdot\,,0)
\,+\,h_0\,(g\,D_+g^{-1})(\,\cdot\,,0)\,h_0^{-1}\ =\ 0\,,&\label{bde2}\\ \cr
&(g^{-1}D_-g)(\,\cdot\,,\pi)\,+\,h_\pi\,(g\,D_+g^{-1})(\,\cdot\,,\pi)
\,h_\pi^{-1}\ =\ 0\,,&
\label{bde3}
\qqq
where $\,D_t=D_+-D_-$ is the covariant derivative along the boundary.
\vskip 0.3cm

The flat gauge field $\,A\,$ may be gauged away by representing it as 
$\,h^{-1}dh\,$ for $h$ mapping the strip to $G$. Setting 
as in the bulk geometry $\,\tilde g=hgh^{-1}\m$ and, on the boundary 
components, $\,\tilde g_0=h\m g_0h^{-1}$, $\,\tilde h_0=h\m h_0h^{-1}$,
and similarly for  $\m\tilde g_\pi$ amd $\m\tilde h_\pi$,
\,we reduce the bulk equations to (\ref{ceghp}) and the boundary equations 
to
\qq
&(\tilde g^{-1}\da_-\tilde g)(\,\cdot\,,0)
\,+\,n_0\,(\tilde g\,\da_+\tilde g^{-1})(\,\cdot\,,0)\,
n_0^{-1}\ =\ 0\,,&\label{tbde2}\\ \cr
&(\tilde g^{-1}\da_-\tilde g)(\,\cdot\,,\pi)\,+\,n_\pi\,
(\tilde g\,\da_+\tilde g^{-1})(\,\cdot\,,\pi)
\,n_\pi^{-1}\ =\ 0\,,&
\label{tbde3}
\qqq
with $\,\tilde h_0\,$ and $\,\tilde h_\pi\,$ equal, respectively,
to constant elements $\,n_0\in C^H_{\nu_0}\,$ and $\,n_\pi\in 
C^H_{\nu_\pi}$. \,The boundary conditions (\ref{GHbc12})
become:
\qq
\tilde g(t,0)\ =\ \tilde g_0(t)\,n_0^{-1}\,\qquad\quad\tilde g(t,\pi)
\ =\ \tilde g_\pi(t)\,n_\pi^{-1}
\label{rbc}
\qqq
for $\,\tilde g_0\,$ mapping the line into $\,C^G_{\mu_0}\,$ and $\,
\tilde g_\pi\,$ into $\,C^G_{\mu_\pi}$. \,As in the bulk case, 
the phase space $\,\CP^{G/H}_{_{M_0M_\pi}}$
of the $G/H$ coset model with the boundary conditions (\ref{GHbc12})
is composed of the gauge-orbits of the classical solutions.
The latter are in one-to-one correspondence with
the orbits of the triples $\,(\tilde g,n_0,n_\pi)\,$
under the simultaneous conjugation by elements of $H$.
In this parametrization, the symplectic form of the boundary
theory is given by
\qq
\Omega^{G/H}_{_{M_0M_\pi}}\ =\ {_k\over^{4\pi i}}\int\limits_0^\pi
\tr\,\Big[-\delta({\tilde g}^{-1}\da_t\tilde g)\,{\tilde g}^{-1}\delta
\tilde g\ +\ 2({\tilde g}^{-1}\da_+\tilde g)\,({\tilde g}^{-1}\delta
\tilde g)^2\Big](t,x)\,\, dx\,\,\cr
+\ {_k\over^{4\pi i}}\,\Big[\,\omega^G_{\mu_0}(\tilde g_0(t))\,\,-\,\,
\omega^H_{\nu_0}(n_0)\,+\,\,\tr\,\,({\tilde g}_0^{-1}\delta
\tilde g_0)(t)\,n_0^{-1}\delta n_0\Big]\,\,\cr\cr
-\ {_k\over^{4\pi i}}\,\Big[\,\omega^G_{\mu_\pi}(\tilde g_\pi(t))\,-\,
\omega^H_{\nu_\pi}(n_\pi)\,+\,\tr\,\,({\tilde g}_\pi^{-1}\delta
\tilde g_\pi)(t)\,n_\pi^{-1}\delta n_\pi\Big]\,\,
\label{oghb}
\qqq
for any fixed $t$.
\vskip 0.3cm

Similarly as in the WZW case, see (\ref{wzcs}) and (\ref{wzm}),
the two-dimensional fields $\,\tilde g\,$ satisfying the classical 
equations (\ref{tbde2}), (\ref{tbde3}) and the boundary conditions 
(\ref{rbc}) may be rewritten in terms of a one-dimensional field 
$\,g_\ell\,$ as  
\qq
\tilde g(t,x)\ =\ g_\ell(x^+)\,\,m_0\,\m\,g_\ell(-x^-)^{-1}\m\,n_0^{-1}
\ =\ g_\ell(x^+)\,\,m_\pi\,\,g_\ell(2\pi-x^-)^{-1}\m\,n_\pi^{-1}
\label{wzcn}
\qqq
with $\,m_0\in C^G_{\mu_0}$, $\,m_\pi\in C^G_{\mu_\pi}\m$
and $\,\,g_\ell:\NR\rightarrow G\,\,$ satisfying
\qq
E\,\,g_\ell\da_yg_\ell^{-1}
\,=\,0\,\quad{\rm and}\,\quad g_\ell(y+2\pi)\,=\,\rho^{-1}g_\ell(y)\,
\gamma\quad{\rm for}\ \ \,\rho=n_0^{-1}n_\pi\,,
\ \ \gamma=m_0^{-1} m_\pi\,.\hspace{0.4cm} 
\label{wzmn} 
\qqq
Given $\,(\tilde g,n_0,n_\pi)$, the triple $\,(g_\ell,m_0,m_\pi)\,$
is determined up to the right multiplication of $\,g_\ell$ by an
element of $G$ accompanied by the adjoint action of its inverse on
$m_0$ and $m_\pi$. \,The symplectic form $\,\Omega^{G/H}_{_{M_0M_\pi}}$
may be rewritten in terms of the data $\,(g_\ell,m_0,n_0,m_\pi,m_0,
n_\pi)\m$. \,The result is given by formula (\ref{ombd}) in Apendix A.
\,The single energy-momentum component of the boundary $G/H$ model is
$\,\,T=-\m{k\over2}\,\tr\,(g_\ell\m\partial_+ g_\ell^{-1})^2$.
\vskip 0.2cm

\subsection{Boundary $\,G/G\,$ model}

In the boundary topological coset $G/G$ theory, the classical
equations imply that $\tilde g_0$,\ $\tilde h_0$,
$\tilde g_\pi$,\ $\tilde h_\pi\,$ and $\,\tilde g=
\tilde g_0\tilde h_0^{-1}=\tilde g_\pi\tilde h_\pi^{-1}\,$ are all 
constant so that the phase space $\,\,\CP^{G/G}_{_{M_0M_\pi}}\,$ of 
the boundary $G/G$ theory is composed of the orbits
under simultaneous conjugations of the quadruples 
\qq
(\tilde g_0,\tilde h_0,\tilde g_\pi,\tilde h_\pi)\ \in\ 
C_{\mu_0}\times C_{\nu_0}\times C_{\mu_\pi}\times C_{\nu_\pi}\ \subset\ G^4
\qquad{\rm with}\quad\ \tilde g_0\tilde h_0^{-1}\,=\,\tilde g_\pi
\tilde h_\pi^{-1}\,.
\label{qdr}
\qqq
The symplectic form is given by
\qq
\Omega^{G/G}_{_{M_0M_\pi}}&=& 
{_k\over^{4\pi i}}\,\Big[\,\omega_{\mu_0}(\tilde g_0)\,\,-\,
\omega_{\nu_0}(\tilde h_0)\,+\,\,\tr\,\,({\tilde g}_0^{-1}\delta
\tilde g_0)\,\m{\tilde h }_0^{-1}\delta\tilde h_0\m\Big]\,\,\cr\cr
&-&{_k\over^{4\pi i}}\,\Big[\,\omega_{\mu_\pi}(\tilde g_\pi)\,-\,
\omega_{\nu_\pi}(\tilde h_\pi)\,+\,\tr\,\,({\tilde g}_\pi^{-1}\delta
\tilde g_\pi)\,{\tilde h }_\pi^{-1}\delta\tilde h_\pi\Big].
\label{oggb}
\qqq
Using this expression, one may check that conjugation invariant 
functions of $\tilde g$ Poisson commute. Below, we shall find a more 
transparent interpretation for the phase spaces of the two-dimensional 
theories described above, including the last example.
\vskip 0.9cm

\nsection{Canonical structure of the \,CS\, theory}
\vskip 0.2cm

The classical Chern-Simons theory \cite{Schwarz,WittCS} is determined 
by the action functional of 
$i\Ng$-valued 1-forms $\CA$ (gauge fields, connections)
on an oriented 3-manifold $\CM$
\qq
S^{CS}(\CA)\ =\ -{_k\over^{4\pi}}\int\limits_{\CM}
\tr\,\Big[\CA\, d\CA\,+\,{_2\over^3}\,\CA^3\Big]
\label{CSa}
\qqq
that does not require a metric on $\,\CM\,$ for its definition.

\subsection{Case without boundary}

If $\,\CM\,$ has no boundary then, under the $G$-valued gauge 
transformations $\,g:\CM\rightarrow G$
\qq
S^{CS}(g\CA g^{-1}\hspace{-0.08cm}+g\,dg^{-1})\ =\ 
S^{CS}(\CA)\ -\ {_k\over^{4\pi}}\int\limits_\CM g^*\chi\,.
\label{CSgi}
\qqq
In particular, the action is invariant under gauge transformations
homotopic to $1$ and, for integer $k$, $\,\ee^{-S^{CS}(A)}\,$ is invariant 
under all gauge transformations. The classical equations $\delta S^{CS}=0$ 
are well known to require that $\,F(\CA)=0\,$ with the solutions 
corresponding to flat connections. In the cylindrical geometry 
$\,\CM=\NR\times\Sigma\,$ with $\,\da\Sigma=\emptyset$, \,the canonical 
closed 2-form on the space of solutions is degenerate along
the gauge directions. Writing $\,\CA=A+A_0dt\,$ where
$A$ is tangent to $\Sigma$ and $t$ is the coordinate of $\NR$,
we may use the gauge freedom to impose the condition
$A_0=0$. In this gauge, the classical equations reduce to
\qq
\da _tA\ =\ 0\,,\qquad F(A)\ =\ 0
\qqq
with the solutions given by static flat connections
on the surface $\Sigma$. The canonical closed 2-form on the space 
of solutions becomes
\qq
\Omega^{CS}\ =\ {_k\over^{4\pi}}\int\limits_\Sigma \tr\,
(\delta A)^2\,.
\label{sfcs}
\qqq
Its degeneration is given by the static gauge transformations 
$\,A\mapsto gAg^{-1}\hspace{-0.08cm}+g\,dg^{-1}$. \,The phase space 
$\,\CP^{CS}\,$ of the CS theory is then composed of the gauge-orbits 
of flat gauge fields $A$. In other words, $\,\CP^{CS}\,$ 
coincides with the moduli space of flat connections on $\m\Sigma$. 
\,Formula (\ref{sfcs}) defines the canonical symplectic structure 
on $\,\CP^{CS}$. \,Below, we shall need several refinements 
of the above well known scheme.
\vskip 0.2cm

\subsection{Wilson lines}

First of all, the CS theory may be coupled to a Wilson line
$\,\CC\subset\CM\,$ marked with a label $\mu$ belonging
to the Cartan subalgebra of $\Ng$. Let $\,\gamma\,$ be a $G$-valued
map defined on the line $\CC$. In the presence of these data,
the action functional is modified to
\qq
S^{CS}(\CA,\gamma)\,=\, S^{CS}(\CA)\, +\,ik\int\limits_{\CC}
\tr\,\mu\,\gamma^{-1}(d+\CA)\gamma\,=\, 
S^{CS}(g\CA g^{-1}\hspace{-0.08cm}+g\,dg^{-1},\,g\gamma)\,
\label{csaw}
\qqq
for $\,g\,$ homotopic to $\,1$. \,The corresponding classical equations 
read
\qq
F(\CA)\,=\,2\pi i\,\gamma\mu\gamma^{-1}\,\CC\,,
\qquad \Big(\,d(\gamma\mu\gamma^{-1})+[\CA,\gamma\mu\gamma^{-1}]\Big)
\m{\CC}\ =\ 0\,,
\qqq
where $\CC$ is viewed as a singular current. They imply that $\CA$ 
is a flat connection with a singularity on $\CC$.
In the cylindrical geometry $\CM=\NR\times\Sigma$ with the Wilson
line $\NR\times\{\xi\}$ we may still go to the $A_0=0$ gauge
in which the classical equations reduce to
\qq
\da_tA\ =\ 0\,,\qquad \da_t(\gamma\mu\gamma^{-1})\ =\ 0\,,
\qquad F(A)\,=\, 2\pi i\m\,\gamma\mu\gamma^{-1}\,\delta_\xi\,.
\label{csw}
\qqq
The canonical 2-form $\,\Omega^{CS}_\mu$ on the space of
solutions has now the form
\qq
\Omega^{CS}_\mu\ =\ \Omega^{CS}\ -\ ik\,\tr\,\mu(\gamma^{-1}d\gamma)^2
\label{sfw}
\qqq
where the last term is the Kirillov-Kostant
symplectic form on the (co)adjoint orbit $\CO_\mu$ in $\Ng$ passing
through $\mu$. The orbits of pairs $(A,\gamma\mu\gamma^{-1})$ 
solving (\ref{csw}) under the time-independent gauge 
transformations\footnote{In fact, the singular terms in (\ref{csw}) 
require some care. A possible way is to consider only solutions 
of (\ref{csw}) that around $\xi$ are of the form 
$\,A=i\,\gamma\mu\gamma^{-1}\m d\varphi\,$, where $\varphi$ 
denotes the argument of a local complex parameter and to admit 
the gauge transformations that are constant around $\xi$. 
Different choices of local parameters lead then to canonically 
isomorphic phases spaces.} form the phase space 
$\,\CP^{CS}_\mu$ of the theory. $\,\Omega^{CS}_\mu$ defines on 
$\,\CP^{CS}_\mu$ the canonical symplectic structure. Of course,
one may consider the CS theory with several Wilson lines. 
\vskip 0.2cm

\subsection{Boundaries}

If the 3-manifold $\CM$ has a boundary then one needs to impose
boundary conditions on the gauge fields $\CA$. In the cylindrical
geometry $\CM=\NR\times\Sigma$ where $\da\Sigma\not=\emptyset$,
we may require that
\qq
A_0\ =\ 0\qquad {\rm on}\quad\ \NR\times\da\Sigma\,.
\label{csbc}
\qqq
The classical equations are still given by $\,F(\CA)=0\,$ 
and the closed 2-form by (\ref{sfcs}). The only modification
is that the degeneration of the latter is given now by the gauge 
transformations equal to $1$ on $\da\Sigma$. The same remarks pertain 
to the case with time-like Wilson lines.
\vskip 0.2cm

\subsection{Double \,CS\, theory}

The last modification of the CS theory on a 3-manifold $\CM=\NR\times\Sigma$
that we shall need is the double theory \cite{MooSei} with a pair
$(\CA,\CB)$ of the, respectively, group $G$ and group $H\subset G$ 
gauge fields. The action functional of the double 
theory is the difference of the CS actions for group $G$ and $H\m$:
\qq
S^{2CS}(\CA,\CB)\ =\ S^{\CS}(\CA)\ -\ S^{CS}(\CB)\,.
\qqq
On the boundary $\NR\times\da\Sigma$ we shall impose
the boundary conditions
\qq
(1-E)\,A_0=0\,,\qquad E\,A_0\ =\ B_0\,,\qquad E\,A_{\tau}\ =\ B_{\tau}\,,
\label{abbc}
\qqq
where $\,A_{\tau}$ denotes the component of $A$ tangent to $\da\Sigma$. 
The phase space of the double theory $\,\CP^{2CS}$ is composed
of the pairs $(A,B)$ of flat connections on $\Sigma$ satisfying
the last condition of (\ref{abbc}), modulo $G$-valued gauge
transformations of $A$ and $H$-valued ones of $B$ that coincide
on the boundary of $\m\Sigma$. The symplectic form 
\qq
\Omega^{2CS}\ =\ {_k\over^{4\pi}}\int\limits_\Sigma
\tr\,\Big[\,(\delta A)^2\ -\ (\delta B)^2\Big]\,.
\label{odb}
\qqq
Clearly, both gauge fields may be coupled to time-like Wilson lines 
with labels in the Cartan subalgebras of $\Ng$ and $\Nh$, respectively. 
In the particular case
when $H=G$, the double CS theory reduces to the single one
on the space $\,\NR\times\tilde\Sigma\,$ with the double surface 
$\tilde\Sigma=\Sigma\#(-\Sigma)$ obtained by gluing $\m\Sigma\m$ along the 
boundary to its copy with reversed orientation. The phase spaces 
reduce accordingly.
\vskip 0.9cm

\nsection{Symplectic relations between the \,WZW, \,coset and
\,CS\, theories} 
\vskip 0.2cm

The symplectic structure of the phase spaces of the WZW and coset
theories is given by the complicated expressions,  see (\ref{owz}), 
(\ref{GHsf0}), (\ref{GGsf}), (\ref{obd}), (\ref{oghb}), (\ref{oggb}), 
(\ref{GHsf}) and (\ref{ombd}). Although obtained by applying 
the general procedures of \cite{COQG}\cite{GTT}, these expressions 
are far from being transparent. On the other hand, the interpretation 
of the symplectic structure of the phase spaces of the CS theory 
determined by the standard constant symplectic form on the space of 
two-dimensional gauge fields and by the Kirillov-Kostant form on 
the coadjoint orbits, see (\ref{sfcs}), (\ref{sfw}) or (\ref{odb}), 
has a clear interpretation. In the present section, we shall describe 
symplectic isomorphisms between the phase spaces of the WZW and coset 
theories and those of the CS theory, elucidating this way the
canonical structure of the first ones. The existence of such 
isomorphisms for the bulk WZW and coset theories has been known for
long time, see \cite{WittCS}\cite{EMSS}\cite{MooSei}. We only give 
their slightly more explicit realization. The isomorphism of the 
boundary WZW theory phase space with a moduli space of flat connections
on a twice punctured disc has been first described in \cite{GTT}.
It represents another aspect of the relations between the boundary 
conformal theories and the topological three-dimensional theories 
developed in \cite{FFFS1}\cite{FFFS2}. 
\vskip 0.2cm

\subsection{Bulk \,WZW\, model}

The bulk WZW model on $\NR\times S^1$ corresponds to the CS theory on 
$\NR\times\CZ$, where $\,\CZ=\{z\,|\,{1\over2}\leq|z|\leq 1\}$, with 
the boundary condition $A_0=0$. The isomorphism $\,I\,$ between the phase 
spaces $\,\CP^{CS}$ and $\,\CP^{WZ}\,$  of the two theories is defined by 
the formula giving the classical solution of the WZW model 
on $\NR\times S^1$ in terms of a flat connection on $\CZ\m$:
\qq
g(t,x)\ =\ P\ \ee^{\,\,\int\limits_{\ell_{x,t}}A}\,,
\qqq
where $\,P\,\ee^{\,\m\int_\ell\,\cdots}\,$ stands for the path-ordered 
(from left to right) exponential and $\ell_{x,t}$ is an appropriate contour, 
see Fig.\,\,2.

\leavevmode\epsffile[-142 -20 178 185]{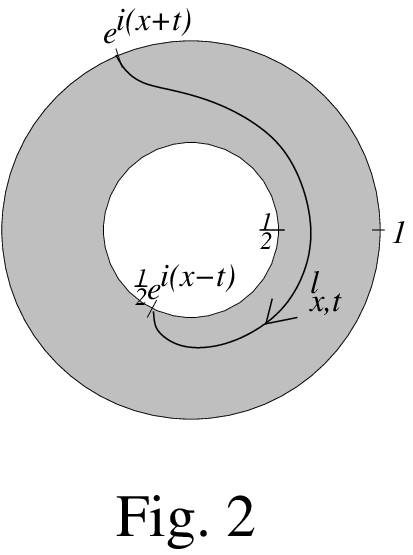}

\noindent In particular, $\,\ell_{x,0}$ is a radial segment from 
$\,\ee^{ix}\,$ to $\,{1\over2}\ee^{ix}\,$ and $\,\ell_{x,t}$ is obtained 
from $\,\ell_{x,0}$ by rotating continuously the beginning of the segment
by angle $t$ and its end by angle $-t$. 
It is not difficult to see that $\,I\,$ is a symplectic isomorphism,
see Appendix B. \,In terms of the CS gauge field $\m A$, 
\m the currents (\ref{curr})
become $\,J_\ell(x^+)=ik\m A_\varphi(\ee^{ix^+})\,$ and
$\,J_r(x^-)=-\m ik\m A_\varphi({_1\over^2}\m\ee^{ix^-})\m$, \,where
$\,A_\varphi\m$ denotes the angular component of $\,A$.
\vskip 0.2cm

\subsection{Bulk $\,G/H\,$ model}

For the bulk $G/H$ coset model, the corresponding CS theory is
the double one on $\NR\times\CZ$. 
Recall that the phase space of the latter is formed by the gauge-orbits
of pairs $\,(A,B)\,$ of, respectively, group $G$ and group $H$ 
flat connections on $\CZ$ whose components tangent to the boundary
are related by $EA_\varphi=B_\varphi$. \,Choose a base point $1\in\CZ$.
We may consider $\,w={1\over i}\ln{z}=\int\limits_1^z{dz\over iz}\,$
as the coordinate on the covering space $\m\tilde\CZ\m$ of $\,\CZ$.
\,Let us inroduce two maps $\,g_{_A}$ and $\,h_{_B}$ from $\m\tilde{\CZ}\m$ 
to $\,G\,$ and $\,H\,$, respectively,
\vskip -0.5cm
\qq
g_{_A}(w)\ =\ P\ \ee^{\,\,\int\limits_{z}^{1}A}\,,\qquad
\qquad h_{_B}(w)\ =\ P\ \ee^{\,\,\int\limits_{z}^{1}B}\,.
\label{tm}
\qqq
Clearly $\,A=g_{_A}dg_{_A}^{-1}\m$ and $\,B=h_{_B}dh_{_B}^{-1}$. 
We shall set
\qq
\tilde g(t,x)\ =\ h_{_B}(x+t)^{-1}\m g_{_A}(x+t)\,
g_{_A}(x-t+w_0)^{-1}\m h_{_B}(x-t+w_0)
\label{tgd}
\qqq
for $\m w_0=i\ln{2}$. Note that $\,\tilde g(t,x+2\pi)=\rho^{-1}
\tilde g(t,x)\,\rho\,$
where 
\qq
\rho\ =\ P\ \ee^{\,\,\int\limits_\CC B}
\label{frho}
\qqq
with $\m\CC\m$ the clock-wise contour around the unit circle 
from $1$ to $1$. It is straightforward to check that 
$\,\tilde g\,$ satisfies the classical equations (\ref{ceghp}) of the
bulk $G/H$ coset model. Under the $G$-valued gauge transformations 
of $A$ and $H$-valued ones of $B$ that agree on the boundary of $\,\CZ$,  
\,the pair $\m(\tilde g,\rho)\m$ undergoes a simultaneous conjugation by 
a fixed element of $H$. We infer that (\ref{tgd}) defines an injective
map $\,I'$ from the phase space $\,\CP^{2CS}\,$ of the double 
CS theory on $\m\NR\times\CZ\,$ to the phase space $\,\CP^{G/H}\,$
of the bulk coset $G/H$ model. Using the parametrization
of the solutions $\,\tilde g\,$ of the coset model by the 
left-right movers 
\qq
g_\ell(y)\ =\ h_{_B}^{-1}(y)\m g_{_A}(y)\,,\qquad
g_r(y)\ =\ h_{_B}(y+w_0)^{-1}g_{_A}(y+w_0)\,,
\label{lrmb}
\qqq
satisfying (\ref{cond}) for $\,\rho\,$ given by (\ref{frho}) and
\vskip -0.5cm
\qq
\gamma\ =\ P\ \ee^{\,\,\int\limits_\CC B}\,,
\label{gtbs}
\qqq
it is easy to see that the map $\,I'$ is also onto. The main result 
is that it defines a symplectic isomorphism, see Appendix B.
\vskip 0.2cm

\subsection{Bulk $\,G/G\,$ model}

In the special case of $H=G$ where the phase space $\,\CP^{2CS}\,$
reduces to that of the gauge-orbits of flat connections on the torus 
represented as the double surface $\,\CZ\#(-\CZ)$, the field 
$\,\tilde g\,$ of (\ref{tgd}) is $(t,x)$-independent and it describes
the parallel transport around the $a$-cycle, see Fig.\,\,3.

\leavevmode\epsffile[-143 -20 157 175]{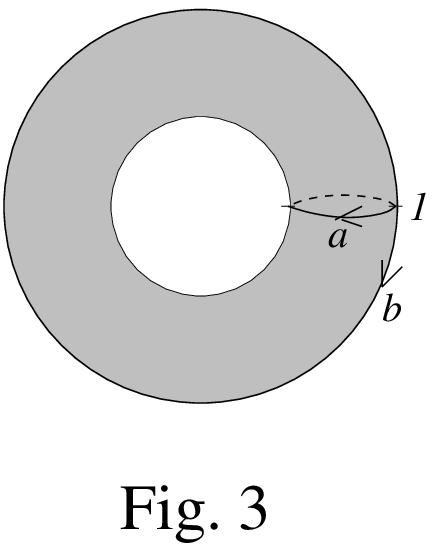}

\noindent Similarly, the monodromy $\,\rho$, \,commuting 
with $\tilde g$, describes the parallel transport around the $b$-cycle. 
Equations (\ref{GGsf}) and (\ref{GGsf1}) express then the symplectic 
form (\ref{sfcs}) in terms of the holonomy of the gauge field and is 
a special case of the result of \cite{AlMal}. Note that 
conjugation-invariant functions of the holonomy around a fixed cycle 
on the torus $\,\CZ\#(-\CZ)$ Poisson-commute.
\vskip 0.2cm

\subsection{Boundary \,WZW\, model}

The case of the boundary WZW model with the boundary conditions
(\ref{bc12}) has been analyzed in \cite{GTT}. The corresponding
CS theory is the one on the solid cylinder $\,\NR\times D\,$ where  
$\,D\,$ is the unit disc in the complex plane, with two time-like 
Wilson lines, say $\,\NR\times\{{_1\over^2}\}\,$ with label $\mu_0$ 
and $\,\NR\times\{-{_1\over^2}\}\,$ with label $-\mu_\pi$, see 
Fig.\,\,4.

\leavevmode\epsffile[-142 -20 163 186]{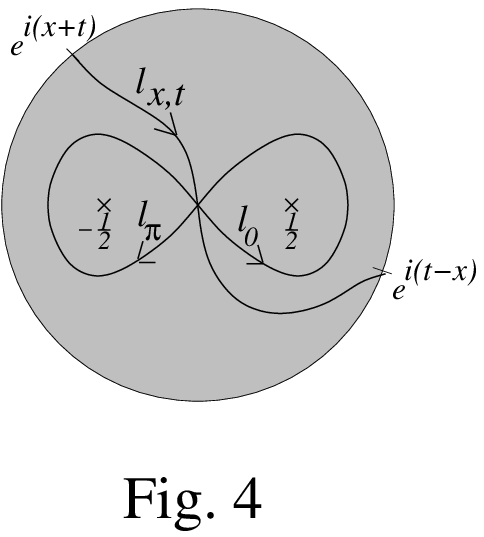}

\noindent Let $\,A\,$ be a flat connection on $\,D\,$ with 
\qq
F(A)\ =\ 2\pi i\m\,\gamma_0\mu_0\gamma_0^{-1}\,\delta_{{_1\over^2}}
\ -\ 2\pi i\m\,\gamma_\pi\mu_\pi\gamma_\pi^{-1}\,\delta_{-{_1\over^2}}\,.
\qqq
Its holonomy around the contour $\,\ell_0\m$ of Fig.\,\,4 lies then 
in the conjugacy class $\,\CC_{\mu_0}$ and around $\,\ell_\pi\m$ in 
$\,C_{\mu_\pi}$. \m To each such connection, we may associate a classical 
solution of the WZW theory on a strip $\,\NR\times[0,\pi]\,$ by 
setting
\qq
g(t,x)\ =\ P\ \ee^{\,\,\int\limits_{\ell_{x,t}}A}\,,
\label{soa}
\qqq
with the contour $\,\ell_{x,t}$ as in Fig.\,\,4. In particular,
for $t=0$, $\,\ell_{x,0}$ goes from $\,\ee^{ix}\m$ to
$\,\ee^{-ix}\m$ crossing once the interval $(-{_1\over^2},{_1\over^2})$.
For other times, $\,\ell_{x,t}$ is obtained by rotating
both ends of  $\,\ell_{x,0}$ by the angle $t$. Note how
the boundary conditions (\ref{bc12}) are assured. The bulk
equations (\ref{cewz}) and the boundary ones (\ref{bewz}) are also
satisfied. The right hand side of (\ref{soa}) is clearly invariant
under the gauge transformations of $A$ equal to $1$ on the boundary
of the disc. The decomposition (\ref{wzcs}) of the solution in terms 
of the one-dimensional field $\m g_\ell\m$ is obtained by setting 
\qq
g_\ell(y)\ =\ P\ \ee^{\,\,\int\limits_{\ell_y}A}\,,
\qquad m_0\ =\ P\ \ee^{\,\,\int\limits_{\ell_0}A}\,,
\qquad m_\pi\ =\ P\ \ee^{\,\,\int\limits_{\ell_\pi}A}\,.
\label{pw}
\qqq
Here for $\m y\in[0,\pi]\m$,  the contour $\,\ell_y$  coincides with
the interval $[\ee^{iy},0]$ and it is deformed continuously for other 
values of $y$. Contours $\,\ell_0$ and $\,\ell_\pi$ are as in Fig.\,\,4.
We obtain this way an isomorphism $\,I_{_{\mu_0\mu_\pi}}$ from the phase 
space $\,\CP^{CS}_{_{\mu_0(-\mu_\pi)}}\,$ of the CS theory on $\NR\times D$ 
with two time-like Wilson lines onto the phase space $\,\CP^{WZW}_{_{
\mu_0\mu_\pi}}$ of the boundary WZW theory. As was explained
in \cite{GTT}, $\,I_{_{\mu_0\mu_\pi}}$ preserves the symplectic
structure. We sketch in Appendix B the idea of the proof.
In \cite{GTT}, this result was used to quantize 
the boundary WZW theory.
\vskip 0.2cm

\subsection{Boundary $\,G/H\,$ model}

Finally, let us consider the coset $G/H$ theory on the strip
$\m\NR\times[0,\pi]$ with the $(M_0,M_1)$ boundary conditions
(\ref{GHbc12}). \,It corresponds to the double CS theory
on $\m\NR\times D\m$ coupled to Wilson lines. The group $G$ gauge field 
is coupled to lines $\m\NR\times\{{_1\over^2}\}\m$ and $\m\NR\times
\{-{_1\over^2}\}\m$ with labels $\mu_0$ and $-\mu_\pi$ and the group $H$
gauge field to the same lines with labels $\nu_0$ and $-\nu_\pi$,
respectively. Let us define
\qq
g_{_A}(y)\ =\ P\ \ee^{\,\,\int\limits_{\ell_y}A}\,,
\qquad h_{_B}(y)\ =\ P\ \ee^{\,\,\int\limits_{\ell_y}B}\,,
\label{gahb}
\qqq
with the contour $\,\ell_y$ as in (\ref{pw}), and
\qq
m_0\ =\ P\ \ee^{\,\,\int\limits_{\ell_0}A}\,,
\quad\ m_\pi\ =\ P\ \ee^{\,\,\int\limits_{\ell_\pi}A}\,,
\quad\ n_0\ =\ P\ \ee^{\,\,\int\limits_{\ell_0}B}\,,
\quad\ n_\pi\ =\ P\ \ee^{\,\,\int\limits_{\ell_\pi}B}.\hspace{0.3cm}
\qqq
The monodromy of $\m g_{_A}$ and $\m h_{_B}$ is given by:
\qq
&&g_{_A}(y+2\pi)\ \,=\m\ g_{_A}(y)\,\gamma\quad\qquad {\rm for}
\qquad\gamma=m_0^{-1}m_\pi\,,\cr\cr
&&h_{_B}(y+2\pi)\ =\ h_{_B}(y)\,\rho\ \ \ \qquad{\rm for}\qquad\rho=
n_0^{-1}n_\pi\,.
\qqq
Setting $\,g_\ell(y)=h_{_B}(y)^{-1}g_{_A}(y)\,$ we obtain a one-dimensional
field satisfying (\ref{wzmn}) and describing via (\ref{wzcn})
a classical solution $\tilde g(t,x)$ of the boundary $G/H$ coset theory
with the $(M_0,M_\pi)$ boundary conditions. Clearly, $\,\tilde g\,$ is
invariant under the gauge transformations of $A$ and $B$ equal
on the boundary of the disc. We obtain this way an isomorphism
$\,I'_{_{M_0M_\pi}}$ between the phase space $\,\CP^{2CS}_{_{M_0
(-M_\pi)}}$ of the double CS theory on $\m\NR\times D\m$ with two pairs
of Wilson lines and the phase space $\,\CP^{G/H}_{_{M_0M_\pi}}$
of the boundary coset model. The proof that $\,I'_{_{M_0M_\pi}}$ 
preserves the symplectic structure is similar to the one in the case 
of the boundary WZW model, see Appendix B.
\vskip 0.2cm

\subsection{Boundary $\,G/G\,$ model}

In the special case $H=G$, the phase space of the double CS theory 
reduces to that of the single theory on the 2-sphere 
$\,S^2=D\#(-D)\,$ with
four Wilson lines: $\m\NR\times\{{_1\over^2}\}\m$ and
$\m\NR\times\{-{_1\over^2}\}\m$ in $\m\NR\times D\m$
with labels $\m\mu_0$ and $-\mu_\pi$ and their images in 
$\m\NR\times(-D)\m$ with labels $-\nu_0$ and $\m\nu_\pi$.
The group elements and $\,\tilde h_0\in C_{\nu_0}$, 
$\,\tilde h_\pi\in C_{\nu_\pi}\m$ and $\,\tilde g=
\tilde g_0\tilde h_0^{-1}=\tilde g_\pi\tilde h_\pi^{-1}$, see 
(\ref{qdr}), are given by the contour integrals
\qq
\tilde h_0\ =\ P\ \ee^{\,\,\int\limits_{\ell'_0}A}\,,
\qquad \tilde h_\pi\ =\ P\ \ee^{\,\,\int\limits_{\ell'_\pi}A}\,,
\qquad\tilde g\ =\ P\ \ee^{\,\,\int\limits_{\ell}A}\,,
\label{ggci}
\qqq
where $\,\ell'_0$ and $\,\ell'_\pi$ are the copies in $-D$
of $\,\ell_0$ and $\,\ell_\pi$, see Fig.\,\,4, and $\,\ell\,$
is the closed contour as in Fig.\,\,5 starting and ending at
the center of $-D$, with the broken pieces contained in $-D$ 
and the solid ones in $D$. 

\leavevmode\epsffile[-138 -20 167 185]{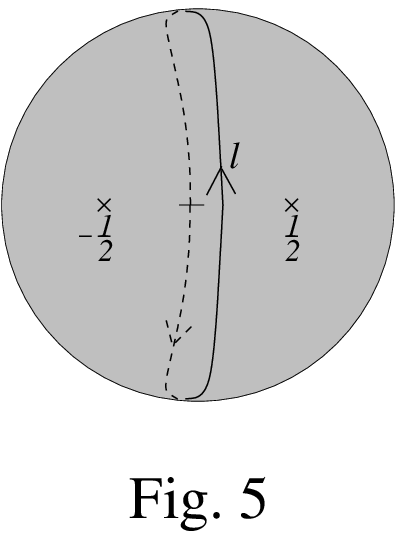}

\noindent The equality of the symplectic form on the moduli space
of flat connections $\,A\,$ on $\,S^2\m$ with four punctures to the 
form of (\ref{oggb}) is essentially again a special case of the result 
of \cite{AlMal}.
\vskip 0.9cm

\nsection{Quantization of the boundary $G/G$ coset model}
\vskip 0.2cm

The description of the canonical structure of the two-dimensional
WZW and coset theories in terms of the moduli spaces of flat
connections on surfaces with boundary may be further reduced
by the ``topological fusion'' to the case of a disc with a single
insertion and of a closed surface with multiple insertions.
This provides a good starting point for the quantization of the 
theory. Indeed, quantization of the moduli spaces on a disc 
with a single puncture is an example of the orbit method in
the representation theory \cite{Kiril}. It gives rise to the highest 
weight representations of the current algebra \cite{AlShat}\cite{EMSS}. 
On the other hand,
quantization of the moduli spaces of flat connections
on closed surfaces leads to the finite-dimensional
spaces of conformal blocks of the WZW theories \cite{WittCS} that
may be also viewed as spaces of invariant 
tensors of the quantum group $\CU_q(\Ng)$ for $q=\ee^{i\pi\over 
k+g^\vee}$ ($g^\vee$ stands for the dual Coxeter number of $G$). 
This was described in detail for the boundary WZW theory with 
$G=SU(2)$ in \cite{GTT}. Here we shall carry out the quantization
program for the boundary coset theory $G/G$. Recall that the canonical 
structure of the $G/G$ model has been described directly in terms of
the moduli spaces of flat connections on closed surfaces
so that no topological fusion will be needed. As a result,
we shall obtain an example of a two-dimensional boundary 
topological field theory, a structure that has recently attracted
some attention \cite{Laza}\cite{Moore}\cite{Moore1}. 
\vskip 0.3cm

Let us start from the well known case of the bulk $G/G$ coset model.
As explained above, the phase space of the theory coincides
with the moduli space of flat connections on the torus
$\m\CZ\#(-\CZ)$. \,The quantization of this moduli space gives 
rise to the space of the conformal blocks $\,\CH\,$ of the the WZW 
theory on the torus\footnote{One 
way to proceed is to use the natural K\"{a}hler structure of
the moduli space given by its identification
with the moduli space of holomorphic vector bundles, 
see e.g.\,\,\cite {CST}}. The space is spanned by the 
affine characters $\chi^k_{_\lambda}$ of level $k$ of the current
algebra $\hat\Ng$ associated to Lie algebra $\Ng$ or by the characters
$\m\chi_{_\lambda}$ of irreducible representations of group $\m G\m$ 
of integrable highest weights $\lambda$ restricted to the 
points\footnote{We view the characters as functions on the 
Cartan algebra identified with functions on the group via the exponential
map} $\,\hat{\zeta}\equiv2\pi{\zeta+\rho_{_W}\over 
k+g^\vee}\,$ for $\m\zeta\m$ running through the integrable weights 
and $\m\rho_{_W}$ standing for the Weyl vector. 
As is well known, the restricted characters induce under pointwise 
multiplication a commutative ring $\CR^k$, the fusion ring of the WZW 
theory. This way, the space $\,\CH\,$ of the conformal blocks on the torus  
becomes a commutative algebra with unity $\,\CH\cong\CR^k\otimes\NC$.
\,The unit element is given by the character $\,\chi_{_0}\equiv1\,$ of the
trivial representation. $\,\CH\,$ may be viewed as the algebra
of functions on the discrete set $\,\{\hat\zeta\,|\,\zeta\ 
{\rm integrable}\m\}=Spec(\CH)\m$. \m The operator of multiplication by 
the restricted character $\m\chi_{_\lambda}$ is the quantizations of the 
function $\,\chi_{_\lambda}(\tilde g)\,$ on the phase space $\,\CP^{G/G}$. 
We may equip $\CH$ with a non-degenerate symmetric bilinear form such 
that\footnote{$\bar\lambda$ denotes the highest weight of the 
representation of $G$ complex conjugate to the one with the highest 
weight $\lambda$.} $\,\langle\m\chi_{_{\lambda}},\m\chi_{_{\lambda'}}\rangle
=\delta_{_{\bar\lambda\m\lambda'}}$ satisfying
\qq
\langle1,\m1\rangle\ =\ 1\,,\qquad\langle a_1a_2,\m a_3\rangle\ 
=\ \langle a_1,\m a_2\m a_3\rangle
\label{scpp}
\qqq
for $\m a_i\in\CH$. 
\,These are the data of a two-dimensional bulk topological field theory
\cite{Atiyah}. Such a theory assigns to each compact, not-necessarily 
connected, oriented surface $\Sigma$ with the boundary $\da\Sigma=\sqcup 
S^1_n$ a linear functional $\,\CI_{_\Sigma}:\mathop{\otimes}\limits_n\CH
\rightarrow\NC$, \,the amplitude of $\m\Sigma$. This assignment 
is supposed to have three properties. First, $\,\CI\,$ is supposed to 
be multiplicative under disjoint products:
\qq
\CI_{_{\Sigma_1\small\sqcup\Sigma_2}}\ =\ \CI_{_{\Sigma_1}}\otimes
\,\CI_{_{\Sigma_2}}\,.\label{djnt}
\qqq	
Second, it should be be covariant with respect to surface 
homeomorphisms preserving orientation. This means that $\,\CI_{_\Sigma}
=\CI_{_\Sigma}\m\pi\,$ for any permutation $\,\pi\,$ of factors  in  
$\,\mathop{\otimes}\limits_n\CH\,$ within the connected components 
of $\m\Sigma$ so that $\,\CI_{_\Sigma}$ depends only on the collection 
of the numbers of boundary circles and handles in each component.
Third, $\,\CI\,$ is required to be consistent with the gluing of boundary 
components:
\qq
\CI_{_{\Sigma_{_{nm}}}}\ =\ \CI_{_{\Sigma}}\m P_{_{nm}}
\label{glue}
\qqq
if $\,\Sigma_{_{nm}}$ is obtained from $\Sigma$ by gluing together the 
$n$-th and the $m$-th boundary components of opposite orientation.
$\,\,P_{_{nm}}=\sum\limits_\gamma id\otimes\dots\otimes
\mathop{p_{_\gamma}}\limits_{\hat{n}}\otimes\dots\otimes\m
\mathop{p'_{_\gamma}}\limits_{\hat{m}}\dots\otimes\m id\,\,$ where 
\qq
P\ =\ \sum\limits_\gamma p_{_\gamma}\otimes\m p'_{_\gamma}\ \in\ 
\CH\otimes\CH
\label{dbf}
\qqq
is the dual of the bilinear form $\,\langle\,\cdot\,,\,\cdot\,\rangle$
(we may take $\,p_{_\gamma}=\chi_{_\lambda}$, $\,p'_{_\gamma}
=\chi_{_{\bar\lambda}}\m$). It is easy to see that if $\,\Sigma\,$
is connected and has $\m g\m$ handles then
\qq
\CI_{_{\Sigma}}\mathop{\otimes}\limits_na_n\ =\ 
\Big\langle\prod\limits_n a_n,\m P^g\Big\rangle\,.
\label{ampl}
\qqq
In particular, the surfaces of Fig.\,\,6 (with the orientation inherited
from the plane) correspond to the amplitudes
\qq
a\mapsto \langle a,1\rangle\,,\qquad a_1\otimes a_2\mapsto\langle a_1,
\m a_2\rangle\,,\qquad a_1\otimes a_2\otimes a_3\mapsto\langle
a_1a_2\m,\,a_3\rangle\,.
\qqq

\leavevmode\epsffile[-54 -20 186 160]{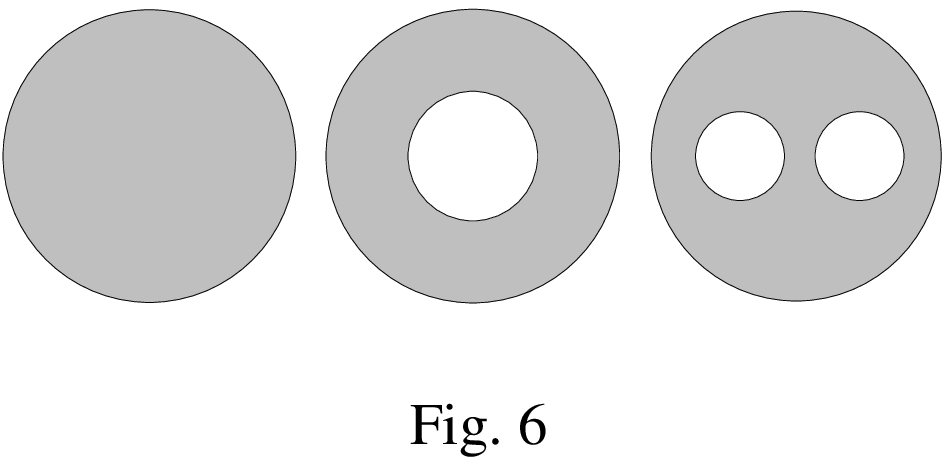}

\noindent They permit to reconstruct the amplitudes for all surfaces.
For the bulk $G/G$ theory and $\,a_n=\chi_{_{\lambda_n}}$, 
expression (\ref{ampl}) 
evaluates to an integer $\,N_{_{(\lambda_n)}}(g)$, \,the Verlinde 
dimension of the space of conformal 
blocks on $\Sigma$ with insertions of the primary fields with labels 
$\lambda_n$. Explicitly \cite{Verl},
\qq
N_{_{(\lambda_n)}}(g)\ =\ \sum\limits_{\zeta}
(S_{_0}^{^\zeta})^{^{2-2g}}\prod\limits_n(S_{_{\lambda_n}}^{^\zeta}/
S_{_0}^{^\zeta})\,,
\qqq
where $\,S^{^\zeta}_{_{\,\lambda}}=S^{^\lambda}_{_{\,\zeta}}=
\overline{S^{^\zeta}_{_{\,\bar\lambda}}}\,$ are the elements of the
matrix giving the modular transformation of the affine characters
$\,\chi^k_{_\lambda}$. For closed surfaces, $\,\CI_{_\Sigma}$ is an 
integer $\,N(g)$. \,It is equal to $1$ for the sphere and to 
$\,{\rm dim}(\CH)$, i.e. the number of integrable weights, for the torus.
In short, the bulk topological $G/G$ coset theory is the theory
of the Verlinde dimensions. They may all be obtained from
the fusion coefficients $\,N_{_{\lambda\eta}}^{^{\,\,\zeta}}=
\m N_{_{\lambda\eta\bar\zeta}}(0)\,$ which define the product
in $\,\CH\m$: $\,\chi_{_\lambda}\chi_{_\eta}
=\sum N_{_{\lambda\eta}}^{^{\,\,\zeta}}\chi_{_\zeta}$. 
\vskip 0.4cm

We would like to extend this structure to the case of the boundary $G/G$ 
coset theory with the $(M_0,M_\pi)$ boundary conditions where 
$\,M_0=(\mu_0,\nu_0)\,$ and $\,M_\pi=(\mu_\pi,\nu_\pi)$. \,Recall
that we have identified the phase space $\,\CP^{G/G}_{_{M_0M_1}}$
of this theory with the moduli space $\,\CP^{2CS}_{_{M_0(-M_\pi)}}$ 
of flat connections on $\,S^2=D\#(-D)\,$ with 
four punctures labeled by $\m\mu_0$, $-\nu_0$, $-\mu_\pi$ and $\m\nu_\pi$, 
all in the Cartan algebra of $G$. \,For $k$ a positive integer 
and $\mu_0=\lambda_0/k$, $\nu_0=\eta_0/k$, $\mu_\pi=\lambda_\pi/k$, 
$\nu_\pi=\eta_\pi/k$, where $\,\lambda_0,\,\,\eta_0,\,\,\lambda_\pi,
\,\,\eta_\pi\,$ are integrable weights, the phase space 
$\,\CP^{2CS}_{_{M_0(-M_\pi)}}$ gives upon quantization the 
space of conformal blocks of the WZW theory on $\m D\#(-D)$ with 
insertions of the primary fields labeled by   
$\lambda_0$ and $\bar\lambda_\pi$ in $D$ and by $\bar\eta_0$ and 
$\eta_\pi$ in $-D$.
We shall denote this space by $\,\CH_{_{L_0L_\pi}}\,$ with
$\, L_0=(\lambda_0,\eta_0)\,$ and $\, L_\pi=(\lambda_\pi,\eta_\pi)$.
\,By the factorization properties of the spaces of conformal blocks, 
\qq
\CH_{_{L_0L_\pi}}\ \cong\ \mathop{\oplus}\limits_{\zeta}\m\,
Hom(\m\CH_{_{\lambda_\pi\bar\eta_\pi\zeta}},\m\CH_{_{\lambda_0
\bar\eta_0\zeta}})
\label{ma}
\qqq
where $\,\CH_{_{\lambda\eta\zeta}}\m$ denotes the space of conformal
blocks on $S^2$ with insertions of three primary fields labeled
by the integrable weights $\lambda$, $\eta$ and $\zeta$.
In particular, $\,\CH_{_{LL}}$ is an associative (in general, 
non-commutative) algebra with unity, a direct sum of matrix
algebras. More generally, there is a natural bilinear product 
$\,\CH_{_{L_1L_2}}\hspace{-0.1cm}\times \CH_{_{L_2L_3}}
\rightarrow\,\CH_{_{L_1L_3}}\,$ 
defined by composition of homomorphisms in each $\zeta$-component.
It gives $\,\CH_{_{L_1L_2}}$ the structure of a left 
$\m\CH_{_{L_1L_1}}$\hspace{-0.1cm}-\m module and of a right
$\,\CH_{_{L_2L_2}}$\hspace{-0.1cm}-\m module. \,It is useful to consider
the direct sum of the boundary spaces $\,\CH_{_b}=\mathop{
\oplus}\limits_{L_1,L_2}\CH_{_{L_1L_2}}$. \,The product in $\,\CH_{_b}$ 
defined by 
\qq
\un{a}\,\un{b}\ =\ =\ \Big(\sum\limits_L a_{_{L_1\hspace{-0.025cm}
L}}b_{_{LL_2}}\Big)\,
\label{bp}
\qqq
for $\,\un{a}=(a_{_{L_1L_2}})\,$ and $\,\un{b}=(b_{_{L_1L_2}})\,$ 
makes $\,\CH_{_b}$ an associative algebra with unity 
$\,\un{1}=(\delta_{_{L_1L_2}})$. 
\vskip 0.3cm

Each space $\,\CH_{_{L_1L_2}}$ is, additionally, a module 
of the commutative algebra $\,\CH\,$ with the character 
$\,\chi_{_\lambda}\in\CR^k\m$ acting diagonally
in the decomposition (\ref{ma}) as the multiplication by 
$\,\chi_{_\lambda}(\hat\zeta)$. This action quantizes the classical 
observables $\,\chi_{_\lambda}(\tilde g)\m$, \,where, in the CS
description, $\,\tilde g\,$ is is the holonomy around the countour
$\ell$ on Fig.\,\,5, see (\ref{ggci}). \,The induced structure of the
$\CH$-module on $\,\CH_{_b}$ satisfies 
\qq
a\,(\un{b}\,\un{c})\ =\ (a\,\un{b})\,\un{c}\ =\ \un{b}\,(a\,\un{c})
\label{mos}
\qqq
for $\m a\in\CH\m$ and $\m\un{b},\un{c}\in\CH_{_b}$ which is equivalent 
to the statement that $\,a\,\un{b}=(a\,\un{1})\,\un b\,$ and that 
elements $\,a\m\un{1}\,$ are in the center of $\,\CH_{_b}$.
\vskip 0.3cm

We shall equip $\,\CH_{_b}\m$ with a non-degenerate symmetric bilinear
form $\,\langle\,\cdot\,,\,\cdot\,\rangle_{_b}\,$ with the only
non-vanishing matrix elements between subspaces 
with permuted boundary labels, i.e. such that
\qq
\langle\m\un{a}\m,\m\un{b}\m\rangle_{_b}\ =\ \sum\limits_{L_1,L_2}
\langle\,a_{_{L_1L_2}},\,b_{_{L_2L_1}}\rangle_{_b}\,.
\label{bpb}
\qqq
Explicitly, we shall set:
\qq
\langle\, a_{_{L_1L_2}},\,b_{_{L_2L_1}}\rangle_{_b}\ =\ 
\sum\limits_\zeta\m (\pm S^{^\zeta}_{_{\,0}})\ \tr\,[\, a_{_{L_1L_2}}
(\zeta)\,\,a_{_{L_2L_1}}(\zeta)\,]\,,
\label{bfb}
\qqq
where the sign is fixed once for all. It is easy to see that the bilinear 
form $\,\langle\,\cdot\,,\,\cdot\,\rangle_{_b}\,$ satisfies
\qq
\langle\m\un{a}\,\un{b}\m,\,\un{c}\m\rangle_{_b}\ 
=\ \langle\m\un{a}\m,\,\un{b}\,\un{c}\m\rangle_{_b}\,.
\label{cyc}
\qqq
The last relation, together with the symmetry
of the form implies the cyclic symmetry
\qq
\langle\m\un{a}\,\un{b}\m,\,\un{c}\m\rangle_{_b}\ 
=\ \langle\m\un{b}\,\un{c}\m,\,\un{a}\m\rangle_{_b}\ 
=\ \langle\m\un{c}\,\un{a}\m,\,\un{b}\m\rangle_{_b}\,.
\label{cycl}
\qqq
Let 
\qq
P_{_{L_1L_2}}\,=\ \sum\limits_A
p^{^A}_{_{L_1L_2}}\hspace{-0.12cm}\otimes\m p^{^A}_{_{L_2L_1}}
\ \in\ \CH_{_{L_1L_2}}\hspace{-0.1cm}\otimes\m\CH_{_{L_2L_1}}
\label{dub}
\qqq
be the dual of the bilinear form (\ref{bfb}) on $\,\CH_{_{L_1L_2}}
\hspace{-0.08cm}\times\m\CH_{_{L_2L_1}}$. We may take 
\qq
A\,=\,(\zeta,i,j)\,\qquad p^{^A}_{_{L_1L_2}}=\,
(\pm S^{^\zeta}_{_{\,0}})^{^{-{1\over2}}}\,e^i_{_{L_1\zeta}}\,
e^{*j}_{_{L_2\zeta}}\,,\qquad p^{^A}_{_{L_2L_1}}
=\,(\pm S^{^\zeta}_{_{\,0}})^{^{-{1\over2}}}
\,e^j_{_{L_2\zeta}}\,e^{*i}_{_{L_1\zeta}}\,,
\label{pal}
\qqq
where, for $\,L=(\lambda,\eta)$, $\,(e^i_{_{L\zeta}})\,$ is a basis 
of $\,\CH_{_{\lambda\bar\eta\zeta}}$ and $\,(e^{*i}_{_{L\zeta}})\,$
is the dual basis. The bilinear forms on $\,\CH_{_b}$ and on 
$\,\CH\,$ are tied together by the relation
\qq
\sum\limits_A\,\langle\,a_{_{L_1L_1}}\m p^{^A}_{_{L_1L_2}}
b_{_{L_2L_2}},\,p^{^A}_{_{L_2L_1}}\rangle_{_b}
\ =\ \sum\limits_\gamma\,\langle\,a_{_{L_1L_1}},\,p_{_\gamma}\m\un{1}
\,\rangle_{_b}\ \langle\,p'_{_\gamma}\m\un{1}\,,
\,b_{_{L_2L_2}}\,\rangle_{_b}\,.
\label{bftt}
\qqq
Indeed, with the use of  (\ref{pal}), the left hand side may be rewritten 
as
\qq
\sum\limits_\zeta\,\tr\,[\,a_{_{L_1L_1}}(\zeta)]\ \tr\,[\,b_{_{L_2L_2}}
(\zeta)]
\qqq
and the right hand side is
\qq
\sum\limits_{\lambda,\zeta,\zeta'}\,S^{^\zeta}_{_{\,0}}\,\,
\chi_{_\lambda}(\hat\zeta)\ S^{^{\zeta'}}_{_{\,0}}
\,\,\chi_{_{\bar\lambda}}(\hat{\zeta'})\ \m\tr\,[\,a_{_{L_1L_1}}(\zeta)]
\ \m\tr\,[\,b_{_{L_2L_2}}(\zeta')]\,.
\qqq
Note that the sign ambiguity in the definition (\ref{bfb}) of the bilinear 
form on $\,\CH_{_b}$ disappears from both expressions. The equality of the 
two sides is inferred by using the relations
$\,S^{^\zeta}_{_{\,0}}\,\chi_{_\lambda}(\hat\zeta)=S^{^\zeta}_{_{\,\lambda}}$,
$\,S^{^{\zeta'}}_{_{\,0}}\m\chi_{_{\bar\lambda}}(\hat\zeta)=
\overline{S^{^{\zeta'}}_{_{\,\lambda}}}\,$ and the unitarity of the modular 
matrix $\,(S^{^\zeta}_{_\lambda})\m$.
\vskip 0.3cm

We may abstract from the above construction an algebraic structure
\qq
\Big(\,\CH,\,\langle\,\cdot\,,\,\cdot\,\rangle,\,
\CH_{_b},\,\langle\,\cdot\,,\,\cdot\,\rangle_{_b}\m\Big)
\label{data}
\qqq
such that
\vskip 0.2cm

1. \ $\CH\,$ is a finite-dimensional associative commutative algebra 
with unity equipped with the non-degenerate symmetric bilinear form 
$\,\langle\,\cdot\,,\,\cdot \,\rangle\m$,
\vskip 0.2cm

2. \ $\CH_{_b}=\oplus\m\CH_{_{L_1L_2}}\,$ is a finite-dimensional 
associative algebra with unity (in general non-commutative) equipped
with the non-degenerate symmetric bilinear form 
$\,\langle\,\cdot\,,\,\cdot\,\rangle_{_b}\m$, 
\vskip 0.2cm

3. \ $\CH_{_b}\m$ is an $\,\CH$-module with each $\,\CH_{_{L_1L_2}}$ 
being a submodule,
\vskip 0.2cm

4. \ relations (\ref{scpp}), (\ref{bp}), (\ref{bpb}), (\ref{cyc}),
(\ref{mos}) and (\ref{bftt}) hold. 
\vskip 0.3 cm

\noindent Such a structure defines a boundary two-dimensional topological 
field theory \cite{Laza}\cite{Moore}\cite{Moore1}. The amplitudes 
of such a theory correspond\footnote{There are minor differences between 
our formulation and that of the above references, mostly a matter
of convenience. In particular, we consider only boundary orientations
induced from the bulk.} to compact oriented surfaces $\,\Sigma\,$ with 
boundary where the boundary components $\m S^1_n$ may contain distinguished 
closed disjoint subintervals (possibly the whole component) marked with 
labels $L$ of the boundary conditions, see Fig.\,\,7.

\leavevmode\epsffile[-97 -20 143 185]{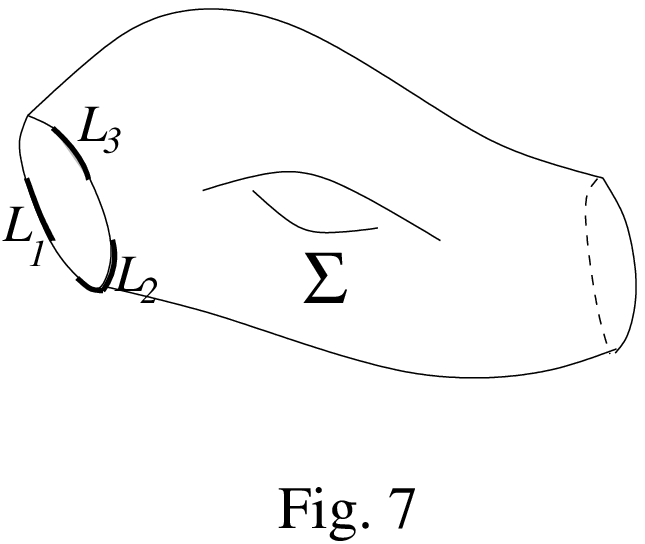}

\noindent Let, for each $\,S^1_n$ with labeled subintervals, $(I_{ns})$ 
be the collection of the remaining subintervals of $S^1_n$.  
The amplitude assigned to such a labeled surface is a linear functional
\qq
\CI_{_\Sigma}:\Big(\mathop{\otimes'}\limits_{n\m}\CH\Big)\otimes
\Big(\mathop{\otimes''}\limits_{n,s\,}\CH_b\Big)\,\rightarrow\,\NC\,,
\label{btft}
\qqq
where the first tensor product is over the boundary components without
labeled subintervals. $\,\CI_{_\Sigma}$ is required to vanish on the
all the components $\,\CH_{_{L_1L_2}}$ of $\,\CH_{_b}$ except those 
with $(L_1,L_2)$ given by the labels of the intervals adjacent to 
$\,I_{ns}$. 

\leavevmode\epsffile[-148 -20 92 160]{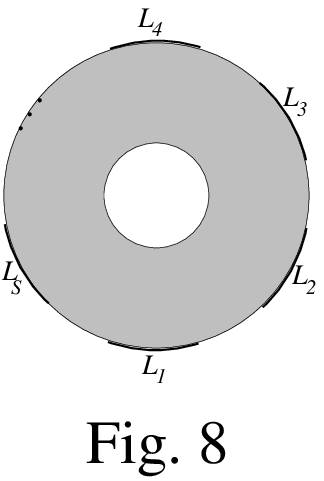}

\noindent For example, the amplitude of the labeled surface of Fig.\,\,8 
is a linear functional on $\,\CH\otimes\CH_{_{L_1L_2}}\hspace{-0.12cm}
\otimes\,\CH_{_{L_2L_3}}\hspace{-0.12cm}\otimes\,\dots\,\otimes\,
\CH_{_{L_{_S}L_{_1}}}$. \,The amplitude assignment $\,\CI\,$ is still 
required to obey (\ref{djnt}) and to be covariant under orientation 
and label preserving homeomorphisms. The latter means that 
$\,\CI_{_\Sigma}=\CI_{_\Sigma}\pi\,$ for cyclic permutations of boundary 
intervals and their labels within boundary circles. The amplitudes 
depend this way on the collections of boundary labels with the cyclic 
order within each boundary circle (including the empty collection).
\,The consistency with gluing (\ref{glue}) is now generalized 
to include the gluing 
along two unlabeled intervals of opposite orientation and permuted 
labels of the adjacent intervals as in Fig.\,\,9. In the latter case 
the dual bilinear form $\,P\in\CH\otimes\CH\,$ should be replaced 
in (\ref{glue}) by the dual form $\,P_{_{L_1L_2}}\hspace{-0.1cm}\in
\CH_{_{L_1L_2}}\hspace{-0.1cm}\otimes\CH_{_{L_2L_1}}$ inserted in the 
appropriate factors of the tensor product $\,\mathop{\otimes''}
\limits_{n,s\,}\CH_b\m$.

\leavevmode\epsffile[-86 -20 154 175]{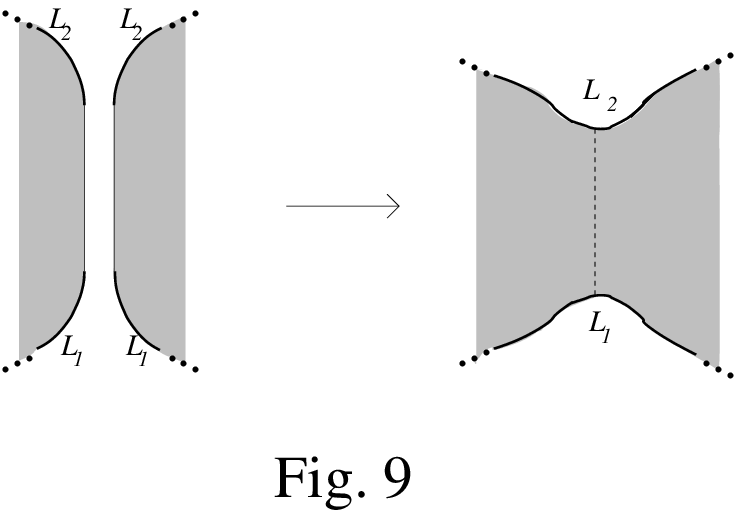}

It is not difficult to construct the amplitudes 
$\,\CI_{_\Sigma}$ from the data (\ref{data}). For completeness,
we shall describe the argument. First, besides the bulk amplitudes 
already discussed, it is enough to know only the amplitudes corresponding 
to labeled surfaces of Fig.\,\,10

\leavevmode\epsffile[-47 -10 193 171]{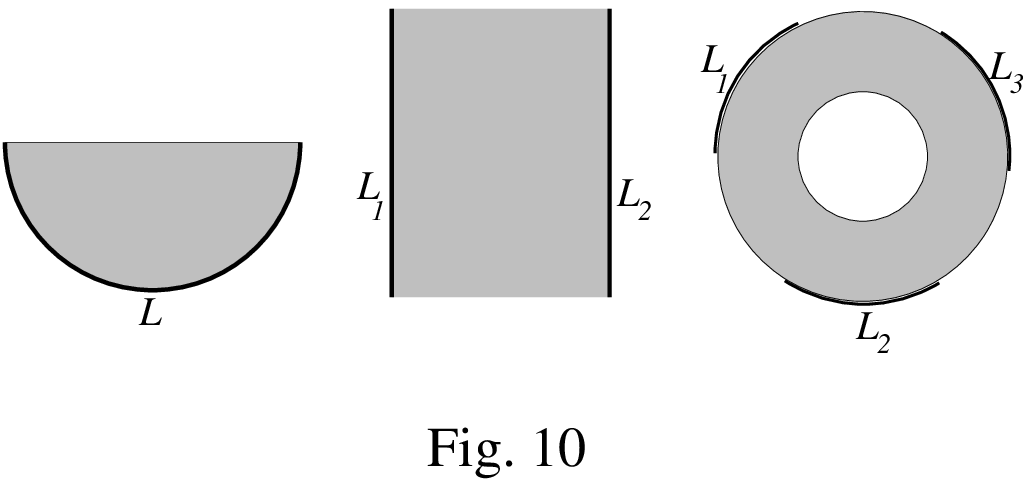}
\qq
&\un a_{_{LL}}\mapsto\ \langle\m a_{_{LL}},\,\un{1}\m\rangle_{_b}\,,
\qquad a_{_{L_1L_2}}\otimes\,b_{_{L_2L_1}}\mapsto\ \,
\langle\m a_{_{L_1L_2}},\, b_{_{L_2L_1}}\rangle_{_b}\,,&\cr\cr
&a\m\,\otimes\, a_{_{L_1L_2}}\otimes\,b{_{L_2L_3}}\otimes
\,c_{_{L_3L_1}}\,\,\mapsto\ \,\langle\,
a_{_{L_1L_2}}b_{_{L_2L_3}},\m\,a\m\,c_{_{L_3L_1}}\rangle_{_b}\,.&
\nonumber
\qqq
Gluing the unlabeled disc to the inner boundary of the annulus
gives the disc with three labeled boundary intervals
and the amplitude
\qq
a_{_{L_1L_2}}\otimes\,b_{_{L_2L_3}}\otimes
\,c_{_{L_3L_1}}\,\,\mapsto\ \,\langle\,
a_{_{L_1L_3}}b_{_{L_3L_2}},\m\,c_{_{L_2L_1}}\rangle_{_b}\,.
\qqq
We may subsequently glue such a disc to the annulus 
as in Fig.\,\,11

\leavevmode\epsffile[-128 -10 112 198]{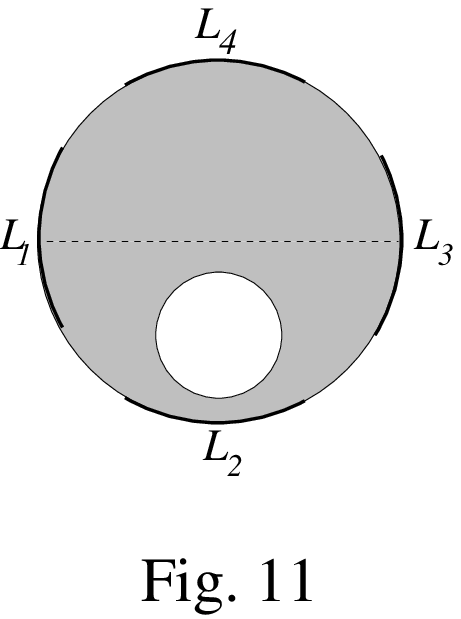}

\noindent to obtain the amplitude
\qq
a\m\,\otimes\,a_{_{L_1L_2}}\otimes\,b_{_{L_2L_3}}\otimes\,c_{_{L_3L_4}}
\otimes\,d_{_{L_4L_1}}\,\mapsto\ \ \sum\limits_{A}
\,\langle\,a_{_{L_1L_2}}b_{_{L_2L_3}},\m\,a\m\,p^{^A}_{_{L_3L_1}}\rangle_{_b}
\,\,\langle\,p^{^A}_{_{L_1L_3}}c_{_{L_3L_4}},\m\,
d_{_{L_4L_1}}\rangle_{_b}\cr\cr
=\ \langle\,a_{_{L_1L_2}}b_{_{L_2L_3}}c_{_{L_3L_4}},\m\,a\m\,d_{_{L_4L_1}}
\rangle_{_b}\,,\hspace{5cm}
\qqq
where the last equality follows from the trivial identity
\qq
\sum\limits_A\,\langle\,a_{_{L_1L_2}}p^{^A}_{_{L_2L_1}}\rangle_{_b}
\ \langle\,p^{^A}_{_{L_1L_2}}b_{_{L_2L_1}}\rangle_{_b}\ =\ 
\langle\,a_{_{L_1L_2}},\m\,b_{_{L_2L_1}}\rangle_{_b}\,.
\label{trid}
\qqq
One obtains similarly the amplitudes of the general annuli 
of Fig.\,\,8. They are given by the linear functionals 
\qq
a\m\,\otimes\,a^{^1}_{_{L_1L_2}}\otimes\,\cdots\,\otimes
\,a^{^S}_{_{L_{_S}L_1}}\ \,\mapsto\,\ \ \langle\,
a^{^1}_{_{L_1L_2}}\,\cdots\,a^{^{S-1}}_{_{L_{_{S-1}}L_{_{S}}}},\,a\,\m
a^{^S}_{_{L_{_S}L_1}}\rangle_{_b}
\label{mbo}
\qqq
\leavevmode\epsffile[-108 -20 132 165]{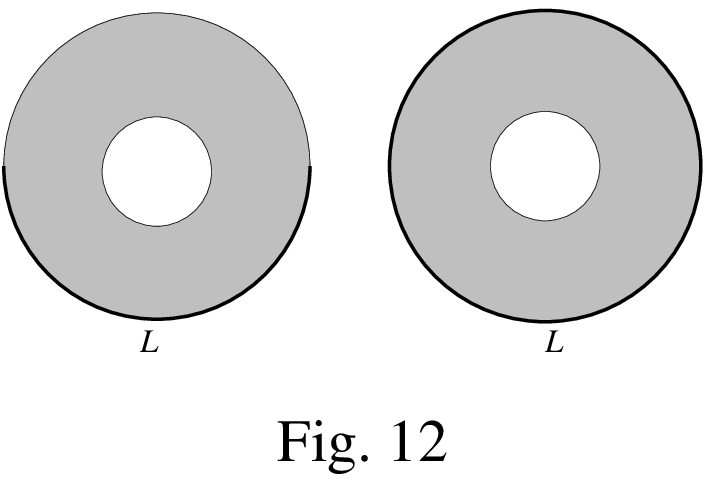}

\noindent invariant under the cyclic permutations of $\,a^{^1}_{_{L_1L_2}}
\cdots a^{^S}_{_{L_{_S}L_1}}\,$ due to (\ref{mos}) and (\ref{cycl}). 
\,The formula extends to the cases with $S=1$ and $S=0$ 
corresponding to the surfaces depicted in Fig.\,\,12
if we interpret it as giving the linear maps
$\,\,a\otimes\m a_{_{LL}}\hspace{-0.08cm}\mapsto\langle\, a_{_{LL}},
\,a\,1_{_{LL}}\rangle\,$ and 
$\,\,a\mapsto\langle\,1_{_{LL}},\,a\,1_{_{LL}}\rangle_{_b}\m$,
\,respectively,
where $\,1_{_{LL}}$ stands for the unity of $\,\CH_{_{LL}}$.
\,For a general surface, we may obtain its amplitude 
by first cutting off the labeled boundary circles around nearby
unlabeled ones as in Fig.\,\,13, and then composing the amplitude 
from that of the annuli of Fig.\,\,8 and of the ones for 
the surface with unlabeled boundary.

\leavevmode\epsffile[-74 -20 166 210]{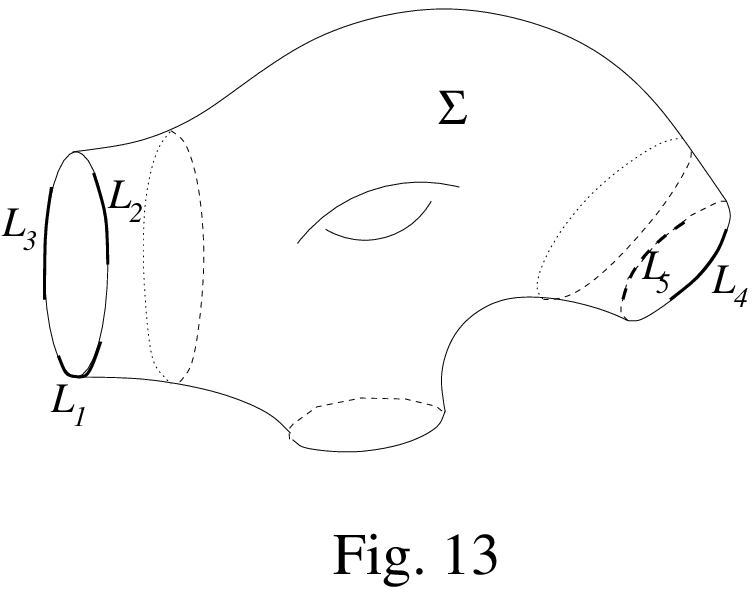}

\noindent It is easy to show that the resulting amplitudes 
are consistent with the gluing of surfaces. For unlabeled surfaces,
this is a well known fact. For surfaces glued along 
two unlabeled boundary intervals in boundary components with
labels, there are two different cases. If the glued intervals 
are in two different boundary components, then the consistency boils 
down to the case of Fig.\,\,14 and it follows with the use of 
(\ref{trid}).

\leavevmode\epsffile[-45 -20 175 175]{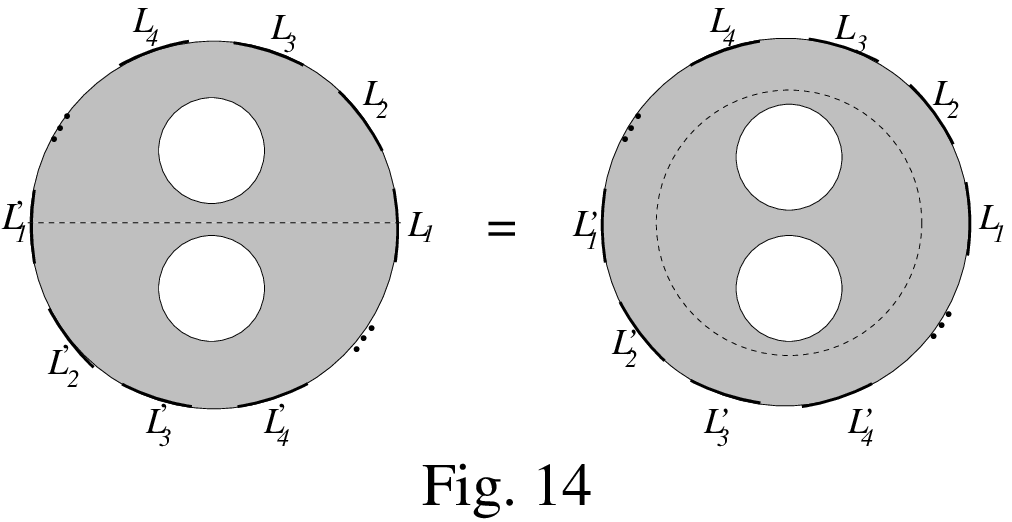}

\noindent The case when one glues two intervals in the same boundary 
components may be similarly reduced to the check that
the gluings of \m Fig.\,\,15 give the same result.

\leavevmode\epsffile[-62 -20 158 170]{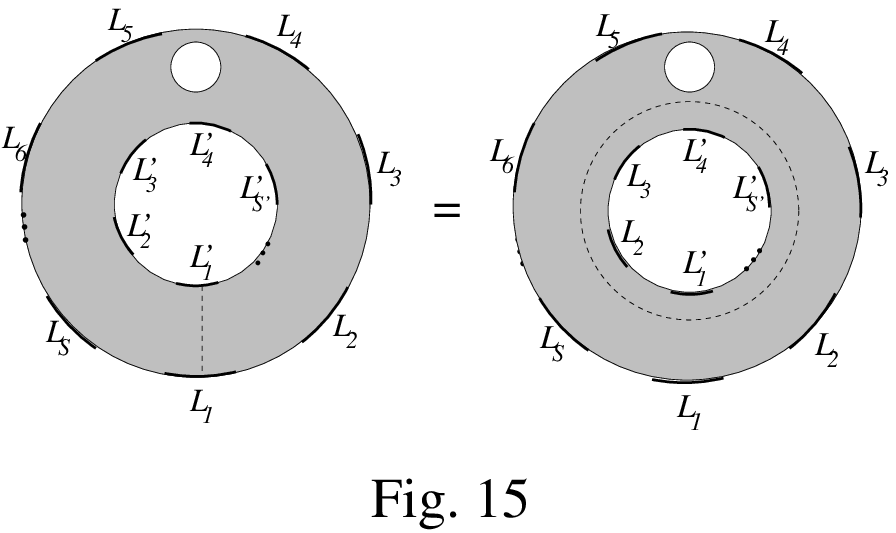}

\noindent The first one leads to the amplitude
\qq
&&a\m\,\otimes\,a^{^1}_{_{L_1L_2}}\otimes\,\cdots\,\otimes
\,a^{^S}_{_{L_{_S}L_1}}
\otimes\,b^{^1}_{_{L'_1L'_2}}\otimes\,\cdots\,\otimes
\,b^{^{S'}}_{_{L'_{_{S'}}L'_1}}\,\,\ \ \,\mapsto\,\,\cr\cr
&&\hspace{2cm}\sum\limits_A\,\langle\,
a^{^1}_{_{L_1L_2}}\,\cdots\,a^{^S}_{_{L_{_S}L_{_1}}}
p^{^A}_{_{L_1L'_1}}
b^{^1}_{_{L'_1L'_2}}\,\cdots\,b^{^{S'}}_{_{L'_{_{S'}}L'_{_1}}},
\,\,a\,\,p^{A}_{_{L'_1L_1}}\m\rangle_{_b}\,.
\label{1tc}
\qqq
The second one results in
\qq
&&a\m\,\otimes\,a^{^1}_{_{L_1L_2}}\otimes\,\cdots\,\otimes
\,a^{^S}_{_{L_{_S}L_1}}
\otimes\,b^{^1}_{_{L'_1L'_2}}\otimes\,\cdots\,\otimes
\,b^{^{S'}}_{_{L'_{_{S'}}L'_1}}\,\,
\ \ \,\mapsto\,\,\cr\cr
&&\hspace{1cm}\sum\limits_{\gamma}
\,\langle\,
a^{^1}_{_{L_1L_2}}\,\cdots\,a^{^{S-1}}_{_{L_{_{S-1}}L_{_S}}},
\,\,p_{_\gamma}\m\,a^{^S}_{_{L_{_S}L_{_1}}}\rangle_{_b}
\ \langle\,b^{^1}_{_{L'_1L'_2}}\,\cdots\,b^{^{S'-1}}_{_{L_{_{S'-1}}
L_{_{S'}}}},\,\,p'_{_{\gamma}}\,\,a\,\,b^{^{S'}}_{_{L_{_{S'}}L'_{_1}}}
\rangle_{_b}\,.\hspace{0.4cm}
\qqq
The equality of both expressions follows from (\ref{bftt}).
\vskip 0.3cm

Conversely, the amplitudes (\ref{btft}) of a two-dimensional boundary
topological field theory define the data (\ref{data}). First, the 
amplitudes of Fig.\,\,6 determine the unity, the bilinear
form and the product in $\,\CH$. \,The commutativity of the latter
follows from the homeomorhism covariance of the amplitudes that allows
to permute the two inner discs of the third surface of Fig.\,\,6.
The associativity of the product in $\,\CH\,$ results from the equality 
of the two ways to glue the amplitudes for the sphere without four 
discs presented in Fig.\,\,16.

\leavevmode\epsffile[-79 -20 141 160]{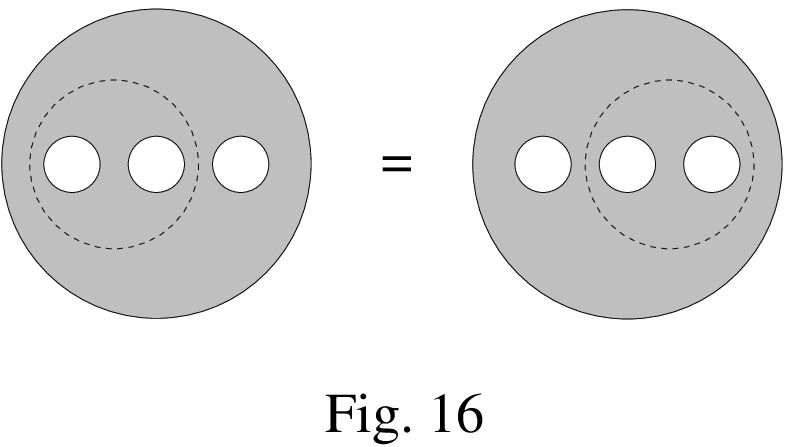}

\noindent The first of the relations (\ref{scpp}) is equivalent 
to the normalization of the amplitude of the sphere $S^2$ to $1$
and the second follows again from the homeomorphism-covariance
of the amplitudes. Similarly, the amplitudes of the first two 
surfaces of Fig.\,\,10 give the unit 
elements $\,1_{_{LL}}\hspace{-0.1cm}\in\CH_{_{LL}}$ and the bilinear 
form pairing $\,\CH_{_{L_1L_2}}$ and $\,\CH_{_{L_2L_1}}$.
\,The amplitude of the third surface applied to $\,1\in\CH\m$, 
\m together with the bilinear form $\,\langle\,\cdot\,,\,\cdot\,
\rangle_{_b}$, \m determine the product $\,\CH_{_{L_1L_2}}
\hspace{-0.1cm}\times \CH_{_{L_2L_3}}\rightarrow\,\CH_{_{L_1L_3}}$ 
in such a way that the cyclic invariance (\ref{cycl}) holds. The 
associativity is proved similarly as before by equating two ways
of gluing a disc with four labeled boundary intervals from pairs
of discs with three labeled boundary intervals, see Fig.\,\,17.

\leavevmode\epsffile[-63 -20 157 180]{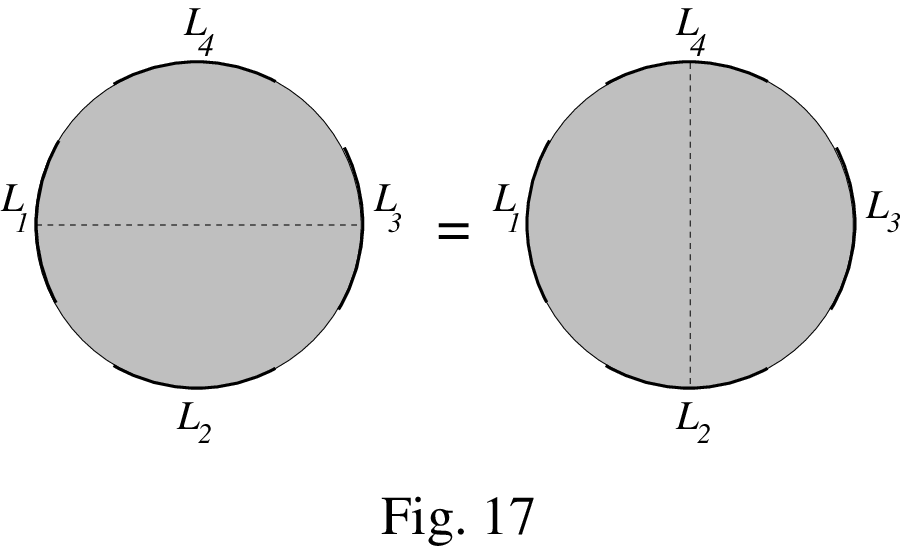}

\noindent The action of the elements of the  bulk space $\,\CH\,$ 
on the boundary space $\,\CH_{_b}$ is obtained from the amplitude 
of the annulus of Fig.\,\,18

\leavevmode\epsffile[-136 -20 174 195]{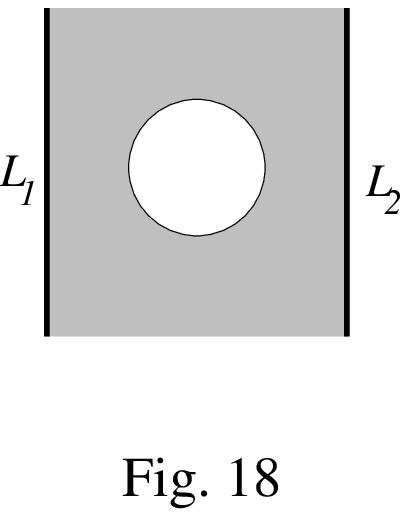}

\noindent with the use of the bilinear form $\,\langle\,\cdot\,,\,\cdot
\,\rangle_{_b}$. \,By definition, this action preserves the subspaces 
$\,\CH_{_{L_1L_2}}\subset\CH_{_b}$. \,The proof that it defines 
a representation of the commutative algebra $\m\,\CH\,$ in 
$\,\CH_{_b}\m$ follows from Fig.\,\,19.

\leavevmode\epsffile[-67 -20 211 173]{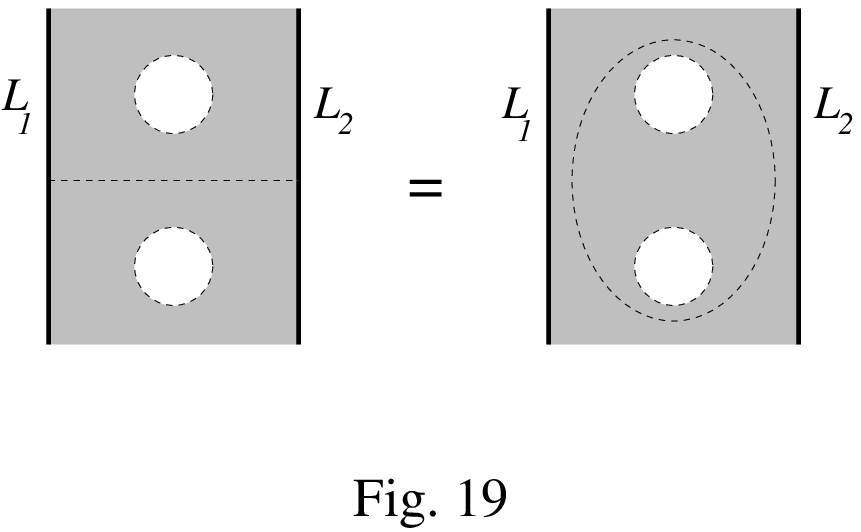}

\noindent Similarly, relations (\ref{mos}) follow from Fig.\,\,20
and (\ref{bftt}) from Fig.\,\,15 with $S=S'=1$.

\leavevmode\epsffile[-37 -20 243 168]{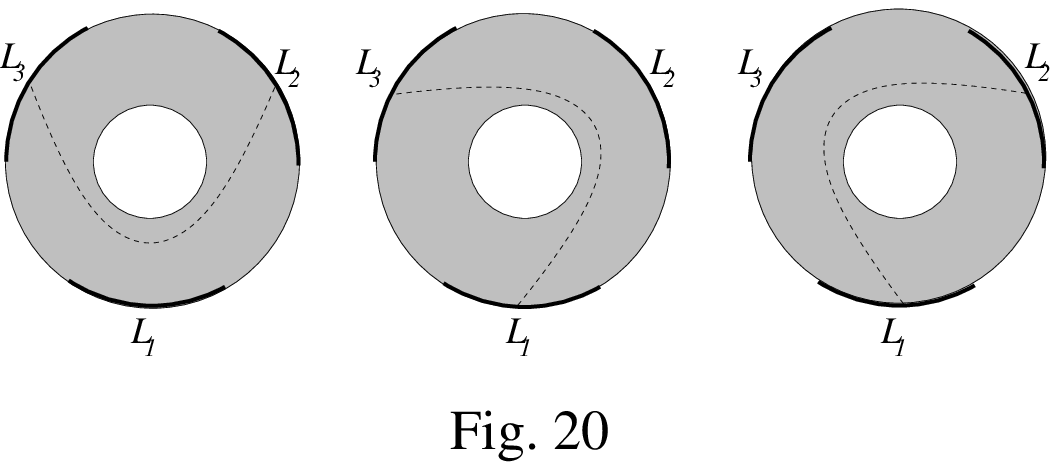}

\noindent We obtain this way the algebraic structure (\ref{data})
possessing all the four properties listed.
\vskip 0.3cm

We shall call a two-dimensional topological field theory unitary
if there exist anti-linear involutions $\,C:\CH\rightarrow\CH\,$
and $\,C_{_b}:\CH_{_b}\rightarrow\CH_{_b}\,$ with 
$\,C_{_b}(\CH_{_{L_1L_2}})=\CH_{_{L_2L_1}}$ such that
the sesqui-linear forms $\,\langle C\,\cdot\,,\,\cdot\,\rangle\,$ and 
$\,\langle C_{_b}\m\cdot\,,\,\cdot\,\rangle_{_b}\,$ define scalar 
products on $\,\CH\,$ and $\,\CH_{_b}$ and that
\vskip 0.2cm

5. \ $C(a\m b)\,=\,(Ca)\m(Cb)\,,\quad \m C_{_b}(\un{a}\,\un{b})\,=\,
(C_{_b}\un{b})\m(C_{_b}\un{a})\,,\quad\m C_{_b}(a\,\un{b})\,=\,
(Ca)\m(C_{_b}\un{b})\,$ for $\,a,b\in\CH\,$ and $\,\un{a},\un{b}
\in\CH_{_b}$.
\vskip 0.2cm

\noindent The last three properties  guarantee that 
\qq
\CI_{_{-\Sigma}}\ = \overline{\CI_{_\Sigma}\,\Big(\mathop{\otimes'}
\limits_{n\m}C\Big)\otimes\Big(\mathop{\otimes''}\limits_{n,s\,}
C_{_b}\Big)}\,,
\label{5or}
\qqq
where $\m-\Sigma\,$ denotes the surface with the reversed orientation
and, conversely, they follow from (\ref{5or}). 
\,For the $\,G/G\,$ theory, one may take for $\,C\,$ the complex 
conjugation of functions of integrable weights and for 
$\,C_{_b}$ the hermitian conjugation of linear transformations
in $\,H_{_{L_0L_\pi}}$, see (\ref{ma}), relative to some scalar 
product in the spaces $\,\CH_{_{\lambda\eta\zeta}}$ of
three-point conformal blocks. One obtains then a unitary
topological field theory provided the sign in (\ref{bfb})
is chosen so that $\,\pm S^{^\zeta}_{_{\,0}}>0$.
\vskip 0.3cm

Recall that, due to (\ref{mos}), the elements $\,a\m1_{_{LL}}$
for $\,a\in\CH\,$ are in the center of $\,\CH_{_{LL}}$. Following 
\cite{Laza}, we shall call the boundary condition $L$ irreducible if
all the elements of the center of $\,\CH_{_{LL}}$ are of this form.
This is the case in the $G/G$ theory. 
\vskip 0.3cm

To each boundary
condition $L$ one may associate a state $\,a_{_L}\in\CH\,$ 
using the amplitude of the second surface of Fig.\,\,12
and the bilinear form on $\,\CH$. \,Explicitly, $\,a_{_L}$
is defined by demanding that
\qq
\langle\,1_{_{LL}},\,a\,1_{_{LL}}\rangle_{_b}\ =\ \langle\m a_{_L},
\,a\m\rangle
\label{ish}
\qqq
for all $\,a\in\CH$. \,We shall call the family of boundary conditions
$\,(\m L\m)\,$ complete if the states $\,(\m a_{_L})\,$ span
$\,\CH$. \,In the $G/G$ theory, for $\,L=(\lambda,\eta)\m$,
\qq
a_{_L}(\hat\zeta)\ =\ N_{_{\lambda
\bar\eta}}^{^{\,\,\zeta}}\,(S_{_0}^{^\zeta})^{^{-1}}
\qqq
and the completeness is easy to see by taking, for example,
the conditions with $L=(\lambda,0)$. On the other hand, the diagonal 
subfamily of boundary conditions corresponding to $\,L=(\lambda,\lambda)\,$
is, in general, not complete since not all integrable weights
appear in the fusion of pairs of complex conjugate weights
(e.g. for $G=SU(2)$, $\,a_{_(j,j)}(\hat{j'})\,$
vanishes for half-integer spins $\m j'$).
\vskip 0.3cm

The bulk topological theories may be perturbed by ``massive''
topological perturbations. For example, in the $SU(2)/SU(2)$
model such perturbations permit to establish a relation with
twisted minimal $N=2$ topological theories. One of the interesting 
open problems for future research is how to extend such relations 
to the case of the boundary $G/G$ theory.

\nappendix{A}
\vskip 0.5cm

When expressed in terms of the left and right movers, the symplectic 
form of the bulk $G/H$ coset theory becomes:
\qq
\Omega^{G/H}\ =\ {_k\over^{4\pi}}\int\limits_0^{2\pi}\tr\,\Big[\,
(g_\ell^{-1}\delta g_\ell)\,\da_y(g_\ell^{-1}\delta g_\ell)\,-\,
(g_r^{-1}\delta g_r)\,\da_y(g_r^{-1}\delta g_r)\,\Big]\,dy\,\,\cr
-\,\,{_k\over^{4\pi}}\,\tr\,\Big[
(\delta\rho)\rho^{-1}\,\Big(
((\delta g_\ell) g_\ell^{-1})(0)\,-\,
((\delta g_r) g_r^{-1})(0)\Big)\,\,\cr\cr
+\,\Big(g_\ell(0)^{-1}(\delta\rho)\rho^{-1}
g_\ell(0)\,-\,g_r(0)^{-1}(\delta\rho)\rho^{-1}g_r(0)\,\,\cr\cr
-\,(g_\ell^{-1}\delta g_\ell)(0)\,
+\,(g_r^{-1}\delta g_r)(0)\Big)(\delta\gamma)\gamma^{-1}
\Big]\,.
\label{GHsf}
\qqq
The expression for the bulk WZW model symplectic form
$\,\Omega^{WZW}\,$ may be obtained from the latter by
setting $\rho$ identically to $1$.
\vskip 0.4cm

Similarly, the expression in terms of the left-mover $\,g_\ell$ 
for the boundary $G/H$ model symplectic form becomes:
\qq
\Omega^{G/H}_{_{M_0M_\pi}}
\ \ =\ \ {_k\over^{4\pi}}\int\limits_0^{2\pi}\tr\,\Big[\,
(g_\ell^{-1}\delta g_\ell)\,\da_y(g_\ell^{-1}\delta g_\ell)\Big]
\,dy\ +\ {_k\over^{4\pi}}\,\tr\,\Big[\,(\delta n_0)
n_0^{-1}\,(\delta n_\pi)n_\pi^{-1}\,\,\cr
-\,\,(\delta m_0)m_0^{-1}\,(\delta m_\pi)m_\pi^{-1}\,
-\,(\delta\rho)\rho^{-1}\,((\delta g_\ell)\,g_\ell^{-1})(0)\,\,\cr\cr
+\,\,(g_\ell^{-1}\delta g_\ell)(0)\,(\delta\gamma)\gamma^{-1}
-\,\,(\delta\rho)\rho^{-1}\,g_\ell(0)\,(\delta\gamma)\gamma^{-1}
g_\ell(0)^{-1}\,\,\cr\cr
+\,\,\omega^G_{\mu_0}(m_0)\,-\,\omega^H_{\nu_0}(n_0)
\,-\,\omega^G_{\mu_\pi}(m_\pi)\,+\,\omega^H_{\nu_\pi}(n_\pi)
\,\Big]\,\,
\label{ombd}
\qqq
and the expression for the boundary WZW model symplectic
form $\,\Omega^{WZW}_{_{\mu_0\mu_\pi}}$ is obtained by setting 
$\rho$, $n_0$ and $n_\pi$ identically to $1$. 
\vskip 0.6cm

\nappendix{B}
\vskip 0.5cm

Our proof of the fact that the isomorphisms $\,I\m$, $\,I'$,
$\,I_{_{\mu_0\mu_\pi}}$ and $\,I'_{_{M_0M_\pi}}$ 
between the WZW and $G/H$ phase spaces and the CS ones preserve 
the symplectic structure is based on a direct calculation of the 
form $\,\int\limits_\Sigma\tr\,(\delta A)^2$ and 
of its counterpart for the gauge field  $B$, very much in the spirit 
of a similar calculation \cite{AlMal} for closed surfaces. 
Consider first the bulk case that is somewhat simpler. It is enough 
to examine the case of the $G/H$ coset theory which for $H=\{1\}$ 
reduces to the WZW model. With $\,g_{_A}$ given by (\ref{tm}),
we have
\qq
\tr\,(\delta A)^2\ =\ d\m\,\tr\m\left[\m({g}_{_A}^{-1}\delta
g_{_A})\,d\m({g}_{_A}^{-1}\delta g_{_A}\m\right]\,.
\label{dela2}
\qqq
Integrating the last expression over the annulus $\,\CZ\,$ cut 
along the interval $\,[{_1\over^2},1]$, \,see Fig.\,\,21, and using the 
Stokes theorem, we infer that
\qq
{_k\over^{4\pi}}\int\limits_\CZ\tr\,(\delta A)^2\ =\ {_k\over^{4\pi}}
\int\limits_0^{2\pi}\tr\,\Big[\,
(g_{_A}^{-1}\delta g_{_A})(y)\,\partial_y\m(g_{_A}^{-1}\delta g_{_A})(y)
\,\,\cr
-\,(g_{_A}^{-1}\delta g_{_A})(y+w_0)\,\partial_y\m(g_{_A}^{-1}
\delta g_{_A})(y+w_0)\,\Big]\,dy\,\,\cr\cr
-\,{_k\over^{4\pi}}\,\tr\,\left[\m(\delta\gamma)\gamma^{-1}\m
\left((g_{_A}^{-1}\delta g_{_A})(0)\,-\,(g_{_A}^{-1}\delta g_{_A})(w_0)
\right)\m\right]\,,
\qqq
where the second line is the contribution from the integrals along 
the cut. Similar expression holds for $B$, $h_{_B}$ and $\rho$
replacing $A$, $g_{_A}$ and $\gamma$, \,respectively.
Subtracting both formulae, we obtain an expression for
the symplectic form of the double CS theory on $\,\CZ\,$ 
which may be shown to coincide with the right hand side of (\ref{GHsf})
by using (\ref{lrmb}) and the second equality of (\ref{cond}).
\vskip 0.4cm

The case of the boundary $G/H$ coset model may be treated similarly.
We define for $\,z\,$ in the unit disc $\,D\,$ cut along the sub-interval 
$\,[-{_1\over^2},1]\,$ of the real axis 
\qq
\tilde g_{_A}(z)\ =\ P\ \ee^{\,\,\int\limits_z^{+i0}A}\,,
\qquad 
\tilde h_{_B}(z)\ =\ P\ \ee^{\,\,\int\limits_z^{+i0}B}\,.
\qqq
Note that for $\,y\in(0,2\pi)$ we have the equalities  
$\,\tilde g_{_A}(\ee^{iy})=g_{_A}(y)\,$ 
and $\,\tilde h_{_B}(\ee^{iy})=h_{_B}(y)\,$ for $\,g_{_A}$
and $\,h_{_B}$ given by (\ref{gahb}). Similarly as before, 
\qq
\int\limits_D\tr\,(\delta A)^2\ =\ \lim\limits_{\epsilon\to0}
\int\limits_{\partial D_\epsilon}\tr\,\left[\m({\tilde g}_{_A}^{-1}\delta
\tilde g_{_A})\,d\m({\tilde g}_{_A}^{-1}\delta\tilde g_{_A}\m\right]\,,
\qqq
where $\,D_\epsilon\m$ is the cut unit disc without $\epsilon$-discs
around $\,\pm{_1\over^2}$, see Fig.\,\,21.

\leavevmode\epsffile[-71 -20 169 181]{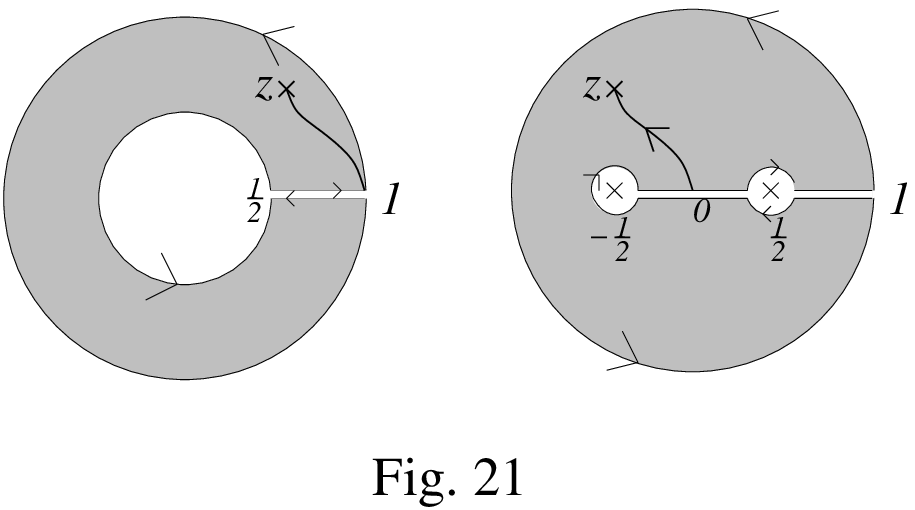}

\noindent A tedious but straightforward calculation results in the 
formula for the symplectic structure of the double CS phase space 
that coincides with equation (\ref{ombd}). We leave the details
to the reader just stressing that a more direct and conceptual
proof of equality between the canonical structures of two-dimensional 
CFT's and three-dimensional topological field theories would be welcome.


\vskip 0.9cm



\begin{thebibliography}{bib}
\bibitem{AlShat}
Alekseev, A., Shatashvili, S.: {\it Quantum groups and WZNW models},
Commun. Math. Phys. {\bf 133} (1990), 353-368 
\bibitem{AlMal}
Alekseev, A., Yu., Malkin, A., Z.: {\it Symplectic structure 
of the moduli space of flat connections on a Riemann surface},
Commun. Math. Phys. {\bf 169} (1995), 99-120 
\bibitem{ASchom}
Alekseev, A. Yu., Schomerus, V.: {\it D-branes in the WZW model},
Phys. Rev. {\bf D 60} (1999), R061901-R061902
\bibitem{Atiyah}
Atiyah, M. F.: {\it Topological quantum field theory},
Publ. IHES {\bf 68} (1989), 175-186
\bibitem{BarRS}
K. Bardak\c{c}i, E. Rabinovici, B. S\"{a}ring, {\it String models 
with}$\ c<1\ ${\it components}, Nucl. Phys. {\bf B 299} (1988), 151-182 
\bibitem{Card} 
Cardy, J.: {\it Boundary conditions, fusion
rules and the Verlinde formula}, Nucl. Phys. {\bf B 324} (1989),
581-596
\bibitem{CatFel}
Cattaneo, A. S., Felder, G.: {\it A path integral approach to 
the Kontsevich quantization formula}, Commun. Math. Phys. {\bf 212}
(2000), 591-611
\bibitem{EMSS}
Elitzur, S., Moore, G., Schwimmer, A., Seiberg, N., 
{\it Remarks on the canonical quantization of the 
Chern-Simons-Witten theory}, Nucl. Phys. {\bf{B\ 326}} (1989), 
104-134 
\bibitem{CST}
Falceto, F., Gaw\c{e}dzki, K.: {\it Chern-Simons States at Genus
One}, Commun. Math. Phys. {\bf 159} (1994), 549-579 
\bibitem{FGbc}
Falceto, F., Gaw\c{e}dzki, K.: {\it Boundary gauged WZW model and
topological Poisson-Lie sigma model}, in preparation
\bibitem{FFFS1}
Felder, G., Fr\"{o}hlich, J., Fuchs, J., Schweigert, C.: {\it Conformal 
boundary conditions and three-dimensional topological field theory},
Phys. Rev. Lett. {\bf 84} (2000), 1659-1662
\bibitem{FFFS2}
Felder, G., Fr\"{o}hlich, J., Fuchs, J., Schweigert, C.: {\it Correlation 
functions and boundary conditions in RCFT and three-dimensional topology},
arXiv:hep-th/9912239
\bibitem{FSSchw}
Fuchs, J., Schellekens, B., Schweigert, C.:
{\it The resolution of field identification 
fixed points in diagonal coset theories}, Nucl.Phys. {\bf B 461} (1996),
371-406
\bibitem{COQG}
Gaw\c{e}dzki, K.: {\it Classical origin of quantum group symmetries 
in Wess-Zumino-Witten conformal field theory}, Commun. Math. Phys. 
{\bf 139} (1991), 201-213
\bibitem{ist}
Gaw\c{e}dzki, K.: {\it Conformal field theory: a case study}.
In: {\it Conformal Field Theory}, Frontiers in Physics 102,
eds. Nutku, Y., Sa\c{c}lioglu, C., Turgut, T., Perseus Publishing,
Cambridge Ma. 2000, pp. 1-55
\bibitem{GK1}
Gaw\c{e}dzki, K., Kupiainen, A.: {\it G/H conformal field theory from
gauged WZW model}, Phys. Lett. {\bf B 215} (1988), 119-123 
\bibitem{GK2}
Gaw\c{e}dzki, K., Kupiainen, A.: {\it Coset construction from functional
integrals}, Nucl. Phys. {\bf B 320} (1989), 625-668 
\bibitem{GTT}
Gaw\c{e}dzki, K., Todorov, I., Tran-Ngoc-Bich, P.: {\it Canonical
quantization of the boundary Wess-Zumino-Witten model}, arXiv:hep-th/0101170
\bibitem{inprep}
Gaw\c{e}dzki, K., Reis, N., Schomerus, V.: in preparation
\bibitem{GKO}
Goddard, P., Kent, A., Olive, D.: {\it Unitary representations of the 
Virasoro and super-Virasoro algebras}, Commun. Math. Phys. {\bf 103} (1986),
105-119 
\bibitem{Hori}
Hori, K.: {\it Global aspects of gauged Wess-Zumino-Witten models},
Commun. Math. Phys. {\bf 182} (1996), 1-32
\bibitem{KPS}
D. Karabali, Q. Park, H. J. Schnitzer, Z. Yang, {\it A GKO construction
based on a path integral formulation of gauged Wess-Zumino-Witten
actions}, Phys. Lett. {\bf B 216}, 307-312 (1989)
\bibitem{Laza}
Lazaroiu, C. I.: {\it On the structure of open-closed
topological field theory in two-dimensions}, Nucl.Phys.
{\bf B 603} (2001), 497-530 
\bibitem{Kiril}
Kirillov, A.: {\it Elements of the Theory of Representations},
Berlin, Heidelberg, New York, Springer 1975
\bibitem{MMS}
Maldacena, J., Moore, G., Seiberg, N.: {\it Geometrical interpretation 
of D-branes in gauged WZW models}, arXiv:hep-th/0105038
\bibitem{Moore}
Moore, G.: {\it Some Comments on Branes, G-flux, and K-theory},
Int.J.Mod.Phys. A16 (2001), 936-944
\bibitem{Moore1}
Moore, G., {\it Santa Barbara lectures}, 
http://online.itp.ucsb.edu/online/mp01
\bibitem{MooSei}
Moore, G., Seiberg, N.: {\it Taming the conformal Zoo},
Phys. Lett. {\bf B 220} (1989), 422-430
\bibitem{PetZub}
Petkova, V., B., Zuber, J.-B.: {\it Conformal boundary conditions 
and what they teach us}, arXiv:hep-th/0103007
\bibitem{Schell}
Schellekens, A. N., Yankielowicz, S.: {\it Field identification
fixed points in the coset construction}, Nucl. Phys. {\bf B 334} (1990),
67-102
\bibitem{Schwarz}
Schwarz, A.: {\it The partition function of degenerate quadratic
functional and Ray-Singer invariants}, Lett. Math. Phys. {\bf 2} (1978),
247-252
\bibitem{Verl}
E. Verlinde, {\it Fusion rules and modular transformations
in 2D conformal field theory}, Nucl. Phys. B {\bf 300} [FS\ {\bf 22}]
(1988), 360-376 
\bibitem{WittWZW}
Witten, E.: {\it Non-abelian bosonization in two dimensions}, Commun. Math.
Phys. {\bf 92} (1984), 455-472 
\bibitem{WittCS}
Witten, E.: {\it Quantum field theory and the Jones polynomial},
Commun. Math. Phys. {\bf 121} (1989), 351-399 

\end{thebibliography}
\end{document}